\documentclass[12pt,draftclsnofoot,onecolumn]{IEEEtran}
\IEEEoverridecommandlockouts

\usepackage[T1]{fontenc}
\usepackage{aecompl}
\usepackage{algorithm}
\usepackage{capt-of}
\usepackage{url}
\usepackage{subfig}
\usepackage{amssymb}
\usepackage{mdwmath}
\usepackage{mdwtab}
\usepackage{arcs}
\usepackage{yhmath}
\usepackage{color}
\usepackage{subeqnarray}
\usepackage{cases}
\usepackage{dsfont}
\usepackage{amsthm}
\usepackage{verbatim}
\usepackage{caption}
\usepackage{color}
\usepackage{diagbox}
\usepackage{setspace}
\usepackage{amsmath}
\usepackage{cite}
\usepackage{amsfonts}
\usepackage{algorithmic}
\usepackage{graphicx}
\usepackage{textcomp}
\usepackage{xcolor}
\usepackage{hyperref}
\usepackage{breakurl}
\usepackage{bm}
\usepackage{steinmetz}
\usepackage{stfloats}
\usepackage{float}
\usepackage{titletoc}
\usepackage{subfig}
\usepackage{subfloat} 
\hypersetup{colorlinks=true,linkcolor=black,citecolor=black,urlcolor=black}
\usepackage[switch,displaymath]{lineno}
\usepackage{bm}
\usepackage{soul}
\usepackage{color}

\def\BibTeX{{\rm B\kern-.05em{\sc i\kern-.025em b}\kern-.08em
    T\kern-.1667em\lower.7ex\hbox{E}\kern-.125emX}}

\begin{document}
\newcommand\degree{^\circ}

\title{\LARGE Graph Neural Networks for Joint  Communication and Sensing Optimization in Vehicular Networks}
\author{Xuefei~Li,~\IEEEmembership{Student Member,~IEEE,}
Mingzhe~Chen,~\IEEEmembership{Member,~IEEE,} 
Yuchen~Liu,~\IEEEmembership{Member,~IEEE,}
Zhilong~Zhang,~\IEEEmembership{Member,~IEEE,}
Danpu~Liu,~\IEEEmembership{Senior Member,~IEEE,}
and~Shiwen~Mao,~\IEEEmembership{Fellow,~IEEE}

\thanks{X. Li, Z. Zhang, and D. Liu are with the Beijing Laboratory of Advanced Information Network, Beijing University of Posts and Telecommunications, Beijing, 100876, China (e-mail: 2013213202@bupt.edu.cn; zhilong.zhang@outlook.com; dpliu@bupt.edu.cn).}

\thanks{M. Chen is with the Department of Electrical and Computer Engineering and Institute for Data Science and Computing, University of Miami, Coral Gables, FL, 33146, USA (e-mail: mingzhe.chen@miami.edu).}
\thanks{Y. Liu is with the Department of Computer Science, North Carolina State University, Raleigh, NC, 27695, USA (e-mail: yuchen.liu@ncsu.edu).}
\thanks{S. Mao is with the Department of Electrical and Computer Engineering, Auburn University, Auburn,
AL, 36849-5201, USA (e-mail: smao@ieee.org).}

\thanks{A preliminary version of this work \cite{Li2023Joint} is accepted by the Proceedings of the 2023 IEEE International Conference on Communications (ICC).}
}

\maketitle

\begin{abstract}
In this paper, the problem of joint communication and sensing is studied in the context of terahertz (THz) vehicular networks. In the studied model, a set of service provider vehicles (SPVs) provide either communication service or sensing service to target vehicles, where it is essential to determine 1) the service mode (i.e., providing either communication or sensing service) for each SPV and 2) the subset of target vehicles that each SPV will serve. The problem is formulated as an optimization problem aiming to maximize the sum of the data rates of the communication target vehicles, while satisfying the sensing service requirements of the sensing target vehicles, by determining the service mode and the target vehicle association for each SPV. To solve this problem, a graph neural network (GNN) based algorithm with a heterogeneous graph representation is proposed. The proposed algorithm enables the central controller to extract each vehicle's graph information related to its location, connection, and communication interference. Using this extracted graph information, a joint service mode selection and target vehicle association strategy is then determined to adapt to the dynamic vehicle topology with various vehicle types (e.g.,  target vehicles and service provider vehicles). Simulation results show that the proposed GNN-based scheme can achieve 93.66\% of the sum rate achieved by the optimal solution, and yield up to 3.16\% and 31.86\% improvements in sum rate, respectively, over a homogeneous GNN-based algorithm and a conventional optimization algorithm without using GNNs.
\end{abstract}

\section{Introduction}
Integration of wireless communication and sensing functionalities on smart vehicles has been regarded as a promising paradigm to improve the safety and efficiency of vehicular networks. The joint design of communication and sensing functionalities can mutually enhance each other by leveraging the unified hardware, spectrum resource, and protocol design \cite{Garcia2021A,Zhang2022Enabling,Zhao2022Radio}. However, the scarce bandwidth of sub-6 GHz bands limits the ability of wireless networks to meet the stringent quality-of-service (QoS) requirements of emerging vehicular applications in terms of delivering high data rates and high-resolution sensing \cite{Christina2021Seven,Chen2019Multi,Wu2021Dynamic,Shojafar2019Energy,Cordeschi2015Reliable}. A promising solution is to use the high-frequency terahertz (THz) bands for its abundant bandwidth. However, deploying THz for joint communication and sensing over vehicular networks faces several challenges, such as severe path loss and extremely directional nature of vehicular links, interference among communication and sensing links, determining which vehicles to provide communication or sensing service, various QoS requirements of communication and sensing, and adaptation to dynamics of vehicle network topology. 

\subsection{Related Work}
 Recently, several works, such as \cite{Liu2022A,Zhong2022Empowering,Petrov2019On,Mu2022NOMA,Liu2020Joint,Zhang2021An,Hieu2020iRDRC,Zhang2021Design}, have studied the problem of resource management for joint communication and sensing. The authors in \cite{Liu2022A} and \cite{Zhong2022Empowering} provided a comprehensive survey of joint communication and sensing systems, and introduced various challenges, problems, and solutions to improve the performance of such systems. The authors in \cite{Petrov2019On} studied the use of the time-domain duplex (TDD) scheme to achieve the fusion of communication and sensing. In \cite{Mu2022NOMA}, the authors optimized the scheduling of communication and sensing signals over different time slots. A time division duplex frame was proposed in \cite{Liu2020Joint} and \cite{Zhang2021An} to determine the communication and sensing mode in each time slot. An adaptive service mode selection algorithm was designed in \cite{Hieu2020iRDRC} to maximize the resource efficiency by selecting the communication and sensing modes based on service requirements. However, the existing works in \cite{Petrov2019On,Mu2022NOMA,Liu2020Joint,Hieu2020iRDRC,Zhang2021An} might introduce mutual interference between communication and sensing systems due to inconsistent operation modes of different vehicles. To address this challenge, in \cite{Zhang2021Design}, the authors analyzed the interference between communication and sensing services so as to optimize the time slot allocation for providing both services to each vehicle. However, these works \cite{Mu2022NOMA,Liu2020Joint,Hieu2020iRDRC,Zhang2021Design,Petrov2019On,Zhang2021An} did not consider the use of THz bands to provide high-rate communication and high-resolution sensing services. Using THz bands can significantly improve both the data rate and sensing resolution, which however, also faces the challenges of higher path loss and attenuation \cite{Chen2021Terahertz}. 

To overcome such limitations, a number of existing works such as \cite{ Boulogeorgos2018Users, Chang2022Joint,Sopin2022User,Xu2021Joint,Wu2021Interference,Shafie2021Coverage,Hossan2022Mobility} studied the use of the THz bands to provide communication service for mobile users. The authors in \cite{Boulogeorgos2018Users} investigated a target vehicle association scheme for ultra-dense THz networks while considering the 
THz channel particularities, the antenna directivity of the base station (BS) and users, as well as their positions and communication service requirements. The authors in \cite{Chang2022Joint} studied the beam alignment problem in THz communications by considering the effect of narrow beamwidth and fast mobility of connected autonomous vehicles. In \cite{Sopin2022User}, the authors analyzed target vehicle association schemes and multi-connectivity strategies for joint THz/millimeter wave (mmWave) deployments. The authors in \cite{Xu2021Joint} optimized spectrum resource allocation for downlink and uplink communications in unmanned aerial vehicle (UAV) assisted THz systems. However, all the above works \cite{Boulogeorgos2018Users,Chang2022Joint,Sopin2022User,Xu2021Joint} simply assumed a constant THz directional antenna gain while ignoring how the main and side lobes affect THz communications.  In \cite{Wu2021Interference}, the authors studied the use of a THz band for communications and modeled the THz antenna gain as a function of beamwidth. However, this work does not consider the radiation of the side lobes of THz antennas. Although the authors in \cite{Shafie2021Coverage} and \cite{ Hossan2022Mobility} studied the directional THz antenna gain of the main lobe and side lobes simultaneously, they did not consider the use of the THz bands to provide sensing service. Therefore, these existing solutions \cite{Wu2021Interference, Boulogeorgos2018Users, Chang2022Joint, Hossan2022Mobility, Shafie2021Coverage,Sopin2022User,Xu2021Joint} cannot be directly applied for vehicle networks that use THz for joint communication and sensing since these two services will interfere each other. Consequently, given the high uncertainty of THz channels, it is critical to manage the vehicle topological information to avoid potential mutual interference when THz bands are used for joint communication and sensing in vehicular networks.  

To manage the vehicle topological information, a number of existing works such as \cite{Sheng2022Graph, Zhao2021Distributed,Qiu2021Topological,Su2022Trajectory, He2020Resource,Hou2021User, Lee2021Graph,Shen2021Graph} studied the use of graph neural network (GNN) to extract vehicle topological information. In \cite{Sheng2022Graph}, a graph convolutional network (GCN) with weighted adjacency matrix was used to capture the spatial features of vehicle topology and describe the intensities of mutual influence between vehicles. The work in \cite{Zhao2021Distributed} trained a GCN to learn the topology related features 
(e.g., vehicle location, vehicle connection, and communication interference) for each user and then solved a link scheduling problem based on the extracted feature vectors. The authors in \cite{Qiu2021Topological} used a topological GCN followed by a sequence-to-sequence framework to predict future traffic flow and density. The authors in \cite{Su2022Trajectory} used a directed GCN to predict the motion trajectories of moving vehicles in complex traffic scenes. 
However, the above works \cite{Zhao2021Distributed,Sheng2022Graph,Qiu2021Topological,Su2022Trajectory} required the information of all vehicles to extract the topology related feature vector for each vehicle, which may not be applied for networks with a dynamic vehicle topology. To make it adaptive to dynamic vehicle topologies, our work herein learns a node representation method with the partial vehicle topological information. The authors in \cite{He2020Resource} and \cite{Hou2021User} proposed to generate topology related feature vectors by sampling and aggregating the information from local neighbors. In \cite{Lee2021Graph}, the authors explored the graph representation methods for link scheduling in device-to-device (D2D) networks, where D2D devices are considered as nodes and the interference among these devices are considered as edges. The authors in 
\cite{Shen2021Graph} selected a fixed number of neighboring vehicles to participate in training to ensure that the input size will not change with the number of vehicles. However, these works \cite{Hou2021User, He2020Resource, Lee2021Graph,Shen2021Graph} only used homogeneous graphs composed of a single type of nodes and edges to represent network devices and their communication links. Hence, all the aforementioned works \cite{Sheng2022Graph, Zhao2021Distributed,Qiu2021Topological,Su2022Trajectory, He2020Resource,Hou2021User, Lee2021Graph,Shen2021Graph} are not suited for extracting topology related feature vectors for joint communication and sensing enabled vehicular networks, since such networks consist of different types of vehicles (e.g., target vehicles and service provider vehicles) and connected links.

\subsection{Contributions}
The main contribution of this work is to design a novel framework that enables service provider vehicles (SPVs) to provide joint communication and sensing services to target vehicles using THz bands. \textit{To the best of our knowledge, this is the first work to study the use of THz for joint communication and sensing in vehicular networks.} Our key contributions include:
\begin{itemize}
    \item We consider the problem of joint communication and sensing over THz vehicular networks. In the studied model, a set of SPVs provide either communication service or sensing service to communication target vehicles or sensing target vehicles, respectively. A central controller determines the service mode (i.e., providing communication or sensing service) for each SPV and the subset of target vehicles that each SPV will serve.
     
    \item We formulate an optimization problem aiming to maximize the sum of the data rates of all communication target vehicles while satisfying the sensing service requirements of sensing target vehicles by jointly determining the service mode (i.e, communication or sensing) and the target vehicle association for each SPV. The THz channel particularities, the directivity of vehicle antennas, as well as the dynamic vehicle topological information and the sensing service requirement are all taken into account in the formulation. 
    
    \item To solve the formulated problem, we propose a novel GNN method that combines GNNs with heterogeneous graphs.
    The proposed algorithm enables the central controller to extract each vehicle's graph information that represents the information related to vehicle location, vehicle connection, and vehicle communication interference. Compared with traditional GNN methods \cite{He2020Resource}, which used a homogeneous graph to represent various types of vehicles, the proposed algorithm adopts a heterogeneous graph with various types of nodes and edges to represent different types of vehicles and their connected links. Using the learned graph information, the probability distribution of each SPV servicing each target vehicle in the corresponding operating mode is obtained. Based on the probability distribution, the non-convex optimization problem can be simplified to a quadratically constrained programming (QCP) problem, which can be solved by the Gurobi tool \cite{Gurobi}.
    
    \item Extensive simulation results show that the number of service provider vehicles, number of target vehicles, vehicle orientation, and dimension of the graph information vector will jointly affect the performance of service mode selection and target vehicle association strategy in THz enabled vehicular networks. In particular, the proposed GNN-based scheme can reach 93.66\% of the sum rate produced by the optimal solution, and yield up to 3.16\% and 31.86\% improvement in the sum rate over a homogeneous graph neural network based algorithm and the conventional optimization algorithm without using GNNs, respectively.
   
\end{itemize}

The remainder of this paper is organized as follows. The system model and the problem formulation are described in Section II. The design of the GNN-based algorithm for service mode selection and target vehicle association is introduced in Section III. In Section IV, numerical results are presented and discussed. Finally, conclusions are drawn in Section V.

\section{System Model and Problem Formulation}
\label{sec:2}
\subsection{Network Model}

\begin{table}[t]
 \renewcommand{\arraystretch}{2}
 \begin{center}
  \caption{List of Main Notation.}
  \label{tab:table1}
  \normalsize
  \resizebox{\textwidth}{!}{
  \begin{tabular}{|c||l|c||l|}   
   \hline
   \makebox[0cm][c]{\textbf{Notation}} & 
   \makebox[11cm][c]{\textbf{Description}}& 
   \makebox[0cm][c]{\textbf{Notation}} & 
   \makebox[10cm][c]{\textbf{Description}}\\
   \hline
   $\mathcal{K}$  & The set of service provider vehicles  &$\mathcal{M}$ & The set of communication target vehicles      \\   \hline
   $\mathcal{N}$ & The set of communication target vehicles &  $S_{km}$ & Received power at vehicle $m$ from vehicle $k$ \\   \hline
   $\theta_{k}$ & Horizontal beamwidth of the antenna for vehicle $k$&  $\varphi_{k}$ & Vertical beamwidth of the antenna for vehicle $k$  \\   \hline
   $G^{\textrm{M}}$ & The antenna gain of the main lobe &  $G^{\textrm{S}}$ & The antenna gain of the side lobes      \\   \hline
   $G^{\textrm{T}}$ & The antenna gain of transmitter &  $G^{\textrm{R}}$ & The antenna gain of receiver \\   \hline
   $L^{\textrm{A}}_{km}$ & Absorption loss between vehicle $k$ and vehicle $m$ &  $\tau(d_{km})$ & Transmittance of the medium \\   \hline
   $L^{\textrm{F}}_{km} $ & Spreading loss between vehicle $k$ and vehicle $m$  & $\phi(f)$ & Overall absorption coefficient of the medium \\   \hline
   $P_{k}$ & Transmit power of vehicle $k$ & $d_{km}$ & Distance between vehicle $k$ and vehicle $m$    \\   \hline
   $I_{km}^{\textrm{C}}(\bm{\alpha},\bm{\beta})$ & Interference to the communication link $k \rightarrow m$  &   $I_{kn}^{\textrm{S}}(\bm{\alpha},\bm{\beta})$ & Interference to the sensing link $k \rightarrow n$ \\   \hline
   $\mathrm{\gamma}_{km}^{\textrm{C}}(\bm{\alpha},\bm{\beta})$ & SINR of the communication link $k \rightarrow m$ &   $\mathrm{\gamma}_{kn}^{\textrm{S}}(\bm{\alpha},\bm{\beta})$ & SINR of the sensing link $k \rightarrow n$ \\   \hline
   $R_{km}^{\textrm{C}}(\bm{\alpha},\bm{\beta})$ &  Data rate of vehicle $k$ transmitting data to vehicle $m$ & $P_k$ & Transmit power of vehicle $k$ \\   \hline
   $B$ &  Spectrum bandwidth  &   $N_0$  &   Johnson-Nyquist noise power \\   \hline
   $\bm{\alpha}$ & Target vehicle association indicator matrix for communication mode & $\bm{\beta}$  &  Target vehicle association indicator matrix for sensing mode \\   \hline
   $\sigma$ & Radar cross section (RCS) & $f$ & Operating frequency \\   \hline
   $\gamma_{\min}$ & The Minimum SINR requirement of the sensing service & $L^{\textrm{S}}_{kn}$ & Spreading loss of the path $k \rightarrow n \rightarrow k$ \\   \hline
  ${\cal{G}}$ & Heterogeneous graph representation &  $\cal{V}$  & Node set of graph ${\cal{G}}$ \\   \hline
   $\cal{E}$ & Edge set of graph ${\cal{G}}$  &  $\cal{O}$ & Node type set of graph ${\cal{G}}$  \\   \hline
   $\cal{R}$ & Edge type set of graph ${\cal{G}}$  &  $\boldsymbol{f}_{v}$ & The feature of vehicle $v$     \\   \hline
   ${g}_{vv^{\prime}}$ & Weight of the edge between vehicle $v$ and vehicle $v^{\prime}$  &  $L$ & Total number of target vehicles  \\   \hline
  $|\cal{K}|$ & The number of service provider vehicles  &  $s_i $ & Sampling size of sampling iteration $i$ \\   \hline
   ${\cal{L}}^1\left(k\right)$ & The set of the first hop vehicles for vehicle $k$ &  ${\cal{L}}^2\left(k\right)$ & The set of the second hop vehicles for vehicle $k$ \\   \hline
   ${\cal{L}}_{R}^1\left(k\right)$ & the subset of the first hop vehicles of vehicle $k$ with the type $R$ edge & $\lambda_{0}$ & The dimension of graph information vector \\   \hline
   $\sigma\left(\cdot\right)$ & Rectified linear unit function (ReLU) &  $\delta\left(\cdot\right)$ & Sigmoid function   \\   \hline $J\left(\boldsymbol{w},\boldsymbol{p},\boldsymbol{b}\right)$ & Binary cross entropy (BCE) loss & $\boldsymbol{w},\boldsymbol{p},\boldsymbol{b}$ & Weight matrices and bias of GNN \\ \hline
   ${\boldsymbol{h}_{k}^{2}}$ & Graph information of vehicle $k$ & $\eta $ & Learning rate \\   \hline
  $\bm{y}_{k}$ & Probability distribution of vehicle $k$ servicing each target vehicle & $z^{l}_{k}$ & The label of vehicle $k$ for class $l$  \\   \hline
  \end{tabular}}
 \end{center}
\end{table}

\begin{figure}[t]
\centering
\includegraphics[width=11cm]{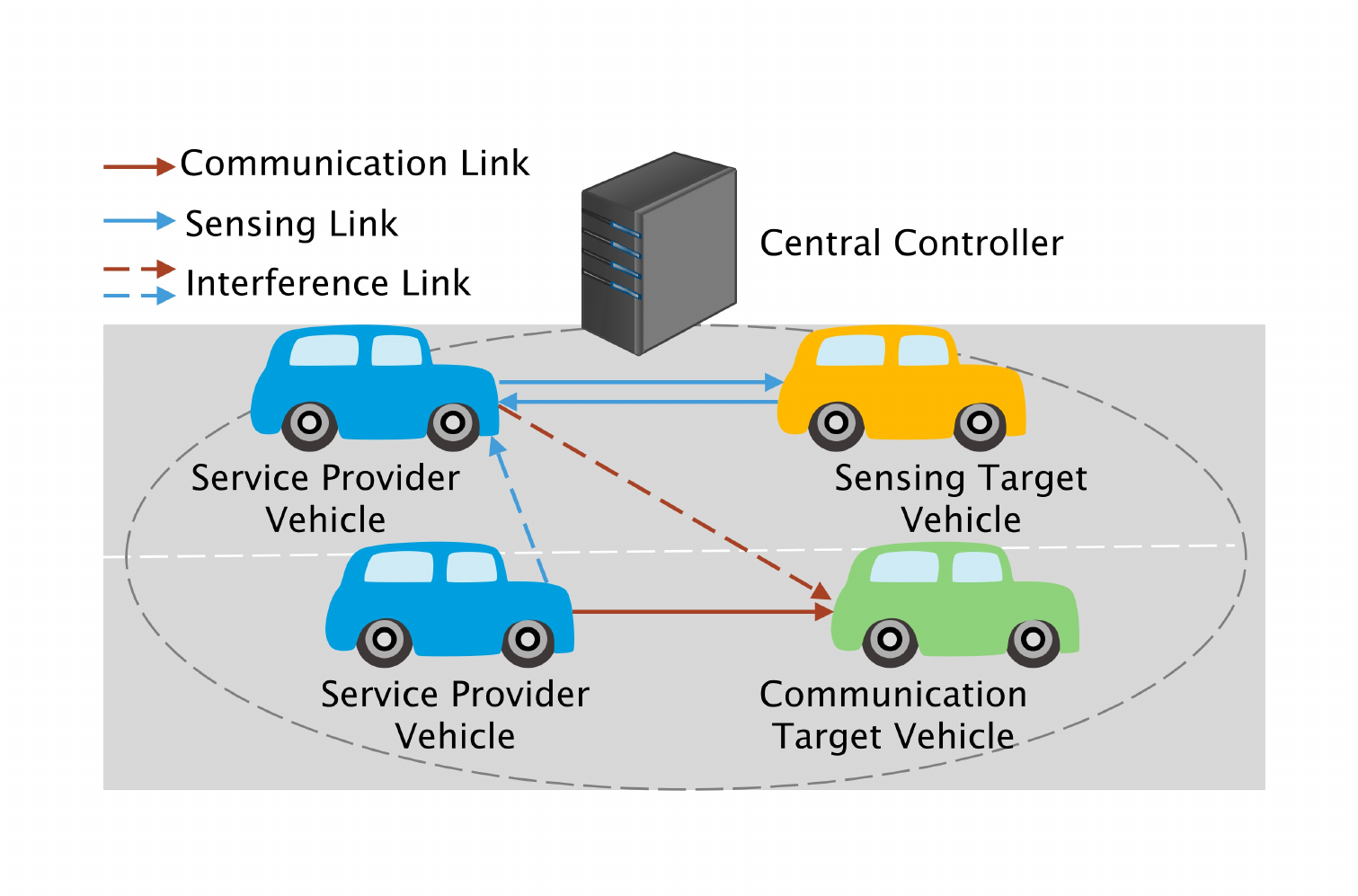}\\
\caption{\normalsize Illustration of the vehicular network model.}
\label{system_model}
\end{figure}

We consider a vehicular network in which  a set of vehicles moving in a region, as shown in Fig.~\ref{system_model}. The vehicles are divided into three categories: service provider vehicles $\mathcal{K}$, communication target vehicles $\mathcal{M}$, and sensing target vehicles $\mathcal{N}$. Each service provider vehicle (SPV) is equipped with both communication and sensing devices, thus can operate in either the communication mode or the sensing mode. When operating in the communication mode, an SPV communicates with the target vehicles through vehicle-to-vehicle (V2V) links. In contrast, an SPV that operates in the sensing mode can sense the location, speed, and direction of the target vehicles for further use (e.g., generate a high-definition map (HD Map)) \cite{Fox2017Multi}. In our model, the locations and density of vehicles vary over time with unknown distributions, and all SPVs use the same THz band to provide communication or sensing services. The main notations used in this paper are summarized in Table \uppercase\expandafter{\romannumeral1}.

\subsubsection{THz Propagation and Antenna Model}
We assume that directional three-dimensional (3D) beams are utilized at the vehicles to compensate for the severe path loss in THz bands. The antenna gains of the main lobe and the side lobes from vehicle $k$ to vehicle $m$ are expressed as\cite{Shafie2021Coverage}
\begin{align}
    & G_{km}^{\textrm{M}}=\frac{4\pi}{(\varepsilon+1) \Omega_{\theta_{k},\varphi_{k}}}, \\
    & G_{km}^{\textrm{S}}=\frac{4\pi \varepsilon}{\left(\varepsilon+1\right)\left(4\pi-\Omega_{\theta_{k},\varphi_{k}}\right)},
\end{align}
where $\Omega_{\theta_{k},\varphi_{k}}=4 \arcsin \left(\tan \left(\frac{\theta_{k}}{2}\right) \tan \left(\frac{\varphi_{k}}{2}\right)\right)$ with $\theta_{k}$ and $\varphi_{k}$ being the horizontal and vertical beamwidths of the antenna for vehicle $k$, and $\varepsilon$ is the ratio of the power concentrated along the side lobes to the power concentrated along the main lobe.

Signal propagation in the THz bands is determined by the molecular absorption loss and the spreading loss. The absorption loss is given by \cite{Wu2021Interference}
\begin{equation}
L^{\textrm{A}}_{km}=\frac{1}{\tau(d_{km})},
\end{equation}
where $d_{km}$ is the distance between vehicle $k$ and vehicle $m$, and $\tau(d_{km}) \approx e^{-\phi(f) d_{km}}$ is the transmittance of the medium following the Beer-Lambert law with  $\phi(f)$ being the overall absorption coefficient of the medium, and $f$ is the operating frequency. Assuming free space propagation, the spreading loss is defined as
\begin{equation}
\begin{aligned}
L^{\textrm{F}}_{km}=\frac{(4\pi f d_{km})^2}{c^2},
\end{aligned}
\end{equation}
where $c$ is the speed of light.

Therefore, the received power at vehicle $m$ from vehicle $k$ can be expressed as 
\begin{equation}
\begin{aligned}
S_{km}=\frac{P_{k}G_{km}^{\textrm{T}} G_{mk}^{\textrm{R}}}{L^{\textrm{A}}_{km} L^{\textrm{F}}_{km}},
\end{aligned}
\end{equation}
where $P_{k}$ is the transmit power of vehicle $k$. $G_{km}^{\textrm{T}}$ and $G_{mk}^{\textrm{R}}$
are the effective antenna gains at vehicle $k$ and vehicle $m$, respectively, corresponding to the link between vehicle $k$ and vehicle $m$ with ${\textrm{T}}\in\left\{{\textrm{M}}, {\textrm{S}}\right\}$ and ${\textrm{R}}\in\left\{{\textrm{M}}, {\textrm{S}}\right\}$, where ${\textrm{M}}$ stands for the main lobe and ${\textrm{S}}$ stands for the side lobes.

\subsubsection{Communication Mode}
The interference to the communication link of vehicle $k$ transmitting to vehicle $m$ is
\begin{equation}\label{eq:int_communication}
\begin{split}
I_{km}^{\textrm{C}}(\bm{\alpha},\bm{\beta})=\sum_{i\in  \mathcal{K}\setminus\{k\}}\sum_{m^{\prime}\in  \mathcal{M}}\frac { \alpha_{im^{\prime}} P_{i} G_{im}^{\textrm{T}} G_{mi}^{\textrm{R}} }{L^{\textrm{A}}_{im}L^{\textrm{F}}_{im}},\\
+\sum_{i\in  \mathcal{K}\setminus\{k\}}\sum_{n^{\prime}\in  \mathcal{N}}\frac {  \beta_{in^{\prime}} P_{i} G_{im}^{\textrm{T}} G_{mi}^{\textrm{R}} }{L^{\textrm{A}}_{im}L^{\textrm{F}}_{im}},
\end{split}
\end{equation}
where $\bm{\alpha}=\left[\bm{\alpha}_{1}, \cdots, \bm{\alpha}_{M}\right]$ with $\bm{\alpha}_{m}=\left[{\alpha}_{1m}, \cdots, {\alpha}_{Km}\right]$, and $\bm{\beta}=\left[\bm{\beta}_{1}, \cdots, \bm{\beta}_{N}\right]$ with $\bm{\beta}_{n}=\left[{\beta}_{1n}, \cdots, {\beta}_{Kn}\right]$.
Here, $\bm{\alpha}$ and $\bm{\beta}$ are the service mode selection and target vehicle association indicator matrices. $\alpha_{im}=1$ indicates that vehicle $i$ is selected to serve vehicle $m$ in the communication mode; otherwise, $\alpha_{im}=0$. Similarly, $\beta_{in}=1$ means that vehicle $i$ is selected to detect vehicle $n$ in the sensing mode; otherwise, $\beta_{in}=0$. In (6), the first term represents the interference from other vehicles that operate in the communication mode, while the second term is the interference from other vehicles that operate in the sensing mode.

The signal-to-interference-plus-noise ratio (SINR) for vehicle $k$ transmitting to vehicle $m$  is
\begin{equation}
\begin{aligned}
\mathrm{\gamma}_{km}^{\textrm{C}}(\bm{\alpha},\bm{\beta})=\frac{ \alpha_{km} S_{km}} {I_{km}^{\textrm{C}}(\bm{\alpha},\bm{\beta})+N_{km}},
\end{aligned}
\end{equation}
where $N_{km}=N_0 +\sum_{i\in  \mathcal{K}\setminus\{k\}}  P_{i} G_{im}^{\textrm{T}} G_{mi}^{\textrm{R}}(1-\tau(d_{im}))/L^{\textrm{F}}_{im}$ with $N_0$ being the Johnson-Nyquist noise power. $N_{km}$ is caused by the thermal agitation of electrons and molecular absorption.



Therefore, the data rate of vehicle $k$ transmitting data to vehicle $m$ is 
\begin{equation}\label{eq:ratev2i}
\begin{aligned}
R_{km}^{\textrm{C}}(\bm{\alpha},\bm{\beta}) = B \log_2 \left(1+\mathrm{\gamma}_{km}^{\textrm{C}}(\bm{\alpha},\bm{\beta})\right), \\
\end{aligned}
\end{equation}
where $B$ denotes the channel bandwidth.

\subsubsection{Sensing Mode}
The interference to vehicle $k$ operating in the sensing mode can be expressed as
\begin{equation}\label{eq:int_sensing}
\begin{split}
I_{kn}^{\textrm{S}}(\bm{\alpha},\bm{\beta})=\sum_{i\in  \mathcal{K}\setminus\{k\}} \sum_{m^{\prime} \in  \mathcal{M}} \frac {  \alpha_{im^{\prime}} P_{i} G_{ik}^{\textrm{T}} G_{ki}^{\textrm{R}} }{L^{\textrm{A}}_{ik}L^{\textrm{F}}_{ik}}  \\
+\sum_{i\in  \mathcal{K}\setminus\{k\}} \sum_{n^{\prime} \in  \mathcal{N}} \frac {  \beta_{in^{\prime}} P_{i} G_{ik}^{\textrm{T}} G_{ki}^{\textrm{R}} }{L^{\textrm{A}}_{ik}L^{\textrm{F}}_{ik}}\\
+ \sum_{i\in  \mathcal{K}\setminus\{k\}} \sum_{n^{\prime} \in  \mathcal{N}}  \frac { \beta_{in^{\prime}} P_{i} G_{in}^{\textrm{T}} G_{nk}^{\textrm{R}} \sigma_{in} c^2}{(4 \pi)^{3} f^2 d_{in}^{2} d_{kn}^{2} L^{\textrm{A}}_{in}L^{\textrm{A}}_{kn}},
\end{split}
\end{equation}
where $\sigma_{in}$ is the target’s radar cross section (RCS) between vehicle $i$ and vehicle $n$.
In (\ref{eq:int_sensing}), the first term represents the interference from other vehicles that operate in the communication mode. The second term represents the interference propagating in the direct path $i \rightarrow k$ from other vehicles that operate in the sensing mode. The third term represents the interference propagating in the scattering path $i \rightarrow n \rightarrow k$ from other vehicles that operate in the sensing mode.

 From (\ref{eq:int_communication}) and (\ref{eq:int_sensing}), we can see that a vehicle that operates in the sensing mode is interfered by other vehicles that operate in the sensing mode from scattering paths, which will not interfere the vehicles that operate in communication mode. This is because the impacts of scattered sensing signals on a communication link is much weaker than that on a sensing link \cite{Zhang2021Design}.

Given (\ref{eq:int_sensing}), the SINR of vehicle $k$ that operates in the sensing mode when sensing vehicle $n$ can be expressed as
\begin{equation}
\mathrm{\gamma}_{kn}^{\textrm{S}}(\bm{\alpha},\bm{\beta})=\frac { \beta_{kn} P_{k} G_{kn}^{\textrm{T}} G_{nk}^{\textrm{R}}  (L^{\textrm{S}}_{kn})^{-1} (L^{\textrm{A}}_{kn})^{-1} }{I_{kn}^{\textrm{S}}(\bm{\alpha},\bm{\beta})+N_{kn}},
\end{equation}
where $L^{\textrm{S}}_{kn}=\frac {(4 \pi)^{3} f^{2} d_{kn}^{4}} {\sigma_{kn} c^2}$ is the spreading loss of the path $k \rightarrow n \rightarrow k$.

\subsection{Problem Formulation}
To maximize the data rates of all communication target vehicles while satisfying the sensing service requirement, an optimization problem is formulated as:

\begin{subequations}\label{eq:litdiff}
\begin{align}
	\mathop{\mbox{max}}_{\bm{\alpha},\bm{\beta}} &  \sum_{k\in \mathcal{K}} \sum_{m\in \mathcal{M}}  R_{km}^{\textrm{C}}(\bm{\alpha},\bm{\beta}) \tag{\ref{eq:litdiff}}\\
	\mbox{s.t.} &\sum_{k\in  \mathcal{K}}\alpha_{km}=1,\alpha_{km} \in \{0,1\},\forall m \in \mathcal{M},\label{eq:a} \\
	&\sum_{k\in  \mathcal{K}}\beta_{kn}=1,\beta _{kn} \in \{0,1\},\forall n \in \mathcal{N},\label{eq:b} \\
	&\sum_{m\in  \mathcal{M}} \alpha_{km}\geq 0, \sum_{n\in  \mathcal{N}}\beta_{kn}\geq 0, \forall k \in \mathcal{K},\label{eq:c}\\
	&\alpha_{km}\beta _{kn}=0, \forall k \in \mathcal{K},\forall m \in \mathcal{M},\forall n \in \mathcal{N},\label{eq:d}\\
	&\sum_{k\in  \mathcal{K}} \mathrm{\gamma}_{kn}^{\textrm{S}}(\bm{\alpha},\bm{\beta}) \geq \gamma_{\min }, \forall n \in \mathcal{N}, \label{eq:e}
\end{align}
\end{subequations}
where $\gamma_{\min}$ is the minimum SINR requirement of the sensing service. In (\ref{eq:litdiff}), constraint (\ref{eq:a}) ensures that a communication target vehicle can only be served by one SPV. Constraint (\ref{eq:b}) ensures that a sensing target vehicle can only be detected by one SPV. Constraint (\ref{eq:c}) indicates that an SPV can serve multiple communication or sensing target vehicles simultaneously. Constraint (\ref{eq:d}) indicates that an SPV can operate in either the communication mode or the sensing mode, but not both simultaneously. Constraint (\ref{eq:e}) is the minimum SINR requirement of the sensing service. 

Problem (\ref{eq:litdiff}) is hard to solve due to the following reasons. First, the objective function is non-convex and hence the complexity of using traditional optimization algorithms will be extremely high. Meanwhile, traditional optimization methods do not consider the dynamic vehicle network topology such as the arrival of new vehicles. Therefore, when the vehicle topology changes, the central controller must execute the optimization algorithm again to re-optimize service mode selection and target vehicle associations. Machine learning (ML) can be developed to learn the relationship between neighboring nodes rather than obtaining a separate feature vector for each vehicle \cite{Chen2021Distributed,Chen2021A}. To solve the above formulated problem, we propose to use GNNs to generate the feature vector for each vehicle. Specifically, it can efficiently obtain the feature vector of a new vehicle without retraining, and then the new service mode selection and target vehicle association strategy can be explicitly determined based on the extracted feature vectors.

\section{Service Mode Selection and Target Vehicle Association based on GNN}
In this section, we introduce a heterogeneous GNN-based algorithm to solve problem (\ref{eq:litdiff}). First, we transform the joint service mode selection and target vehicle association problem (\ref{eq:litdiff}) into a classification problem, where SPVs and target vehicles are considered as samples and classes, respectively. Since each SPV can simultaneously provide communication/sensing service for multiple communication/sensing target vehicles, the corresponding problem naturally becomes a multi-label classification problem, where each sample belongs to a set of classes. We study the use of a heterogeneous GNN-based algorithm to solve this classification problem. Specifically, we introduce the use of a heterogeneous graph to represent the considered system model, and then introduce the components of our designed algorithm. We will also explain the training method for the designed algorithm. 
\begin{figure*}[t]
\centering
\includegraphics[width=16 cm]{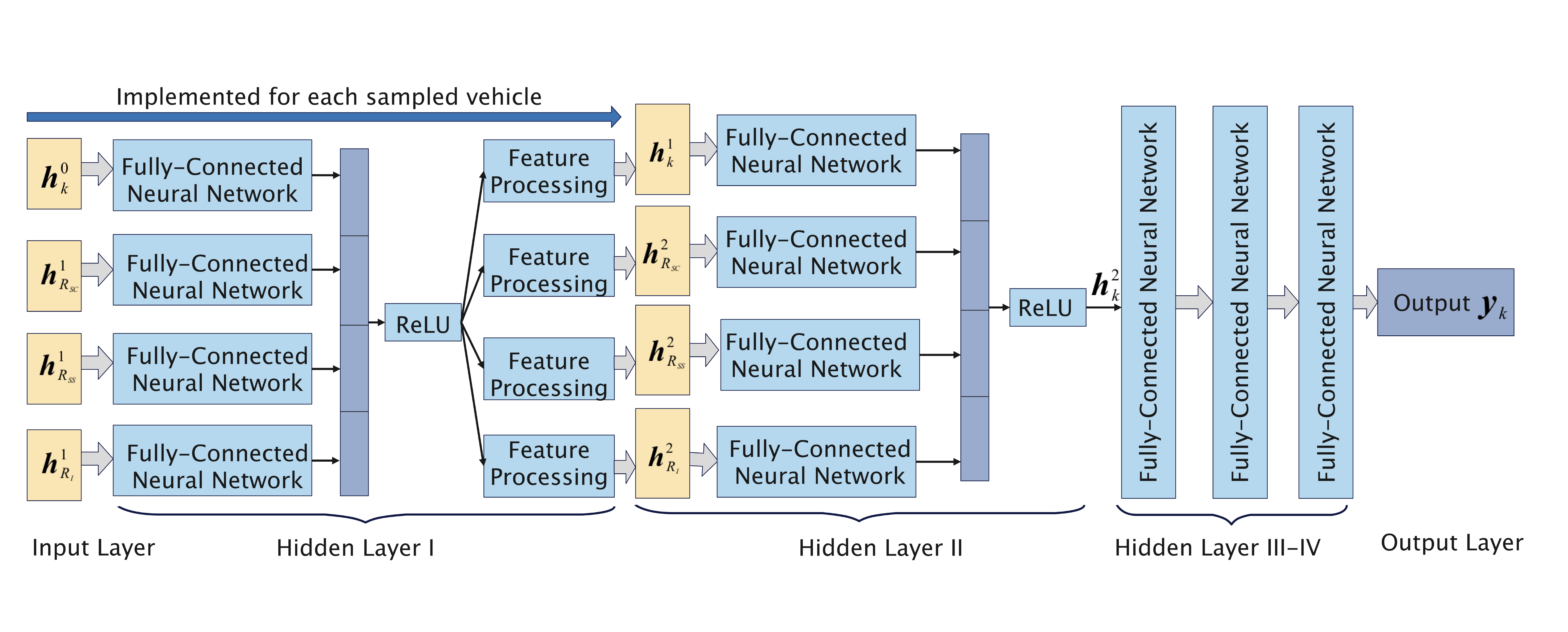}
\caption{\normalsize Structure of the proposed GNN model.}
\label{nn}
\end{figure*}

\subsection{Graph Representation of Vehicular Networks}
\begin{figure}[t]
\centering
\subfloat[Vehicle topology.]{
\begin{minipage}[t]{0.4\linewidth}
\centering
\includegraphics[width=3in]{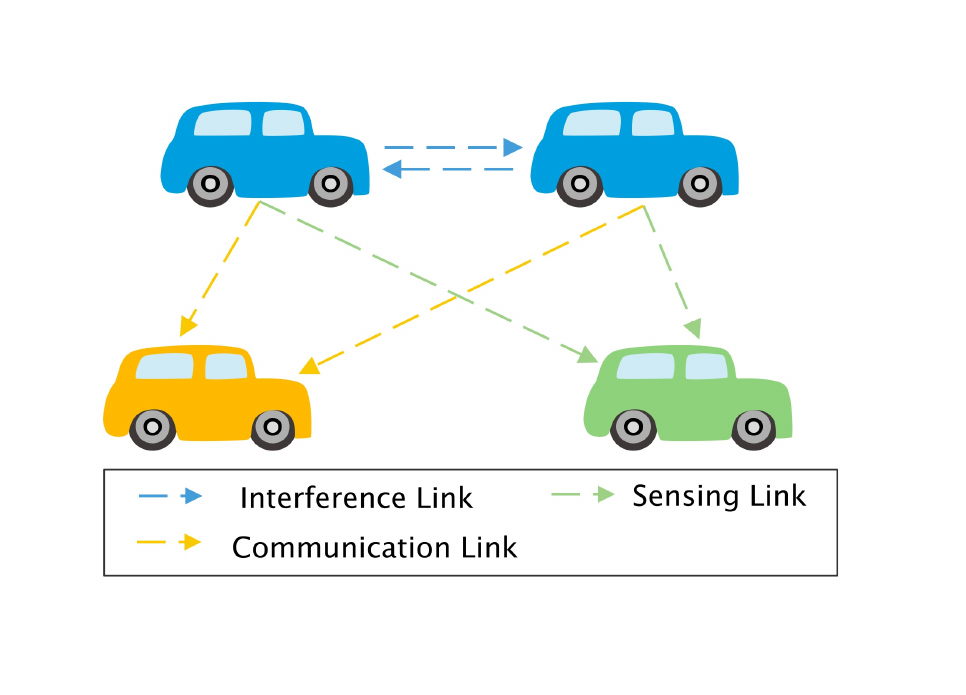}
\label{p2a}
\end{minipage}
}
\subfloat[ Graphical model.]{
\begin{minipage}[t]{0.55\linewidth}
\centering
\includegraphics[width=4.5in]{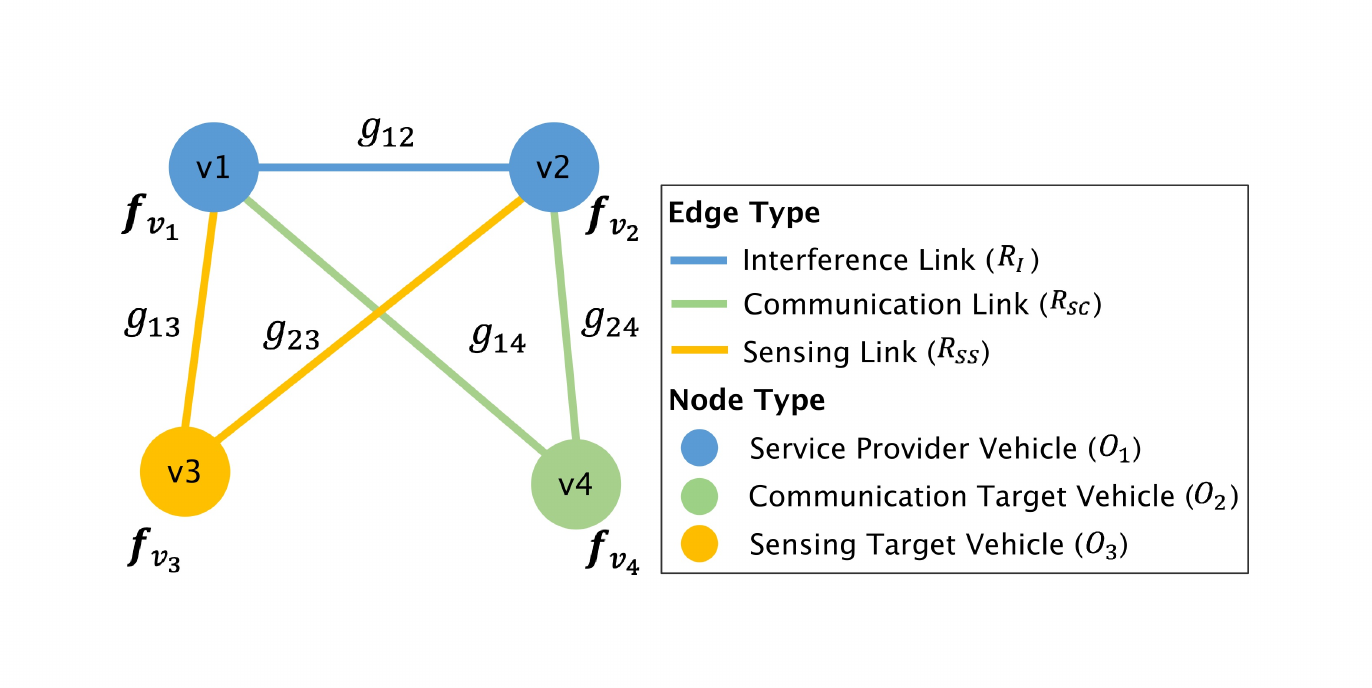}
\label{p2b}
\end{minipage}
}
\centering
\caption{\normalsize Graph representation of the vehicular network.}
\label{Representation}
\end{figure}

We first introduce the use of a heterogeneous graph to represent the considered vehicular network. A heterogeneous graph $\cal{G} = \left(V, E, O, R\right)$ consists of a node set $\cal{V}$, an edge set $\cal{E}$, a node type set $\cal{O}$, and a set $\cal{R}$ that consists of different edge types. We model each vehicle as a node in the graph, and each link between two vehicles as an edge. The nodes can be divide into three categories,  ${\cal{O}}=\{O_1, O_2, O_3\}$,  which correspond to the three types of vehicles. Meanwhile, we consider three types of edges ${\cal{R}}=\{R_\emph{SC}, R_\emph{SS}, R_\emph{I}\}$, where $R_\emph{SC}$ represents the communication link between an SPV and a communication target vehicle, $R_\emph{SS}$ represents the sensing link between an SPV and a sensing target vehicle, and $R_\emph{I}$ represents the interference link between two SPVs. By using the above graph representation, we can convert the vehicular network shown in Fig.~\ref{Representation}\subref{p2a} into graphical model shown in Fig.~\ref{Representation}\subref{p2b}. Specifically, the node feature of each vehicle is $\boldsymbol{f}_{v}=\left[e_{v1},\dots,e_{vM^{\prime}}\right], {\cal{V}} \in \cal{K} \cup \cal{M} \cup \cal{N},{\cal{M^{\prime}}}=\cal{M} \cup \cal{N}$, where $\boldsymbol{f}_{v} \in \mathbb{R}^{ L \times 1}$, $L=\left(|\cal{M}|+|\cal{N}|\right)$ is the total number of target vehicles, and $e_{vm^{\prime}}$ is the number of SPVs within the line-of-sight link between vehicle $v$ and vehicle $m^{\prime}$, as shown in Fig.~\ref{Beam}. The node feature evaluates the potential interference between the current vehicle and each target vehicle. The weight of the edge between vehicle $v$ and vehicle $v^{\prime}$ is ${g}_{vv^{\prime}}=\left(L^{\textrm{A}}_{vv^{\prime}}L^{\textrm{F}}_{vv^{\prime}}\right)^{-1}$ for all $v^{\prime} \in {\cal{V}}\setminus\{v\}$ with $\boldsymbol{g}_{vv^{\prime}} \in \mathbb{R}^{ 1 \times 1}$.

\subsection{Components of the GNN-based Algorithm}
Next, we will introduce the components of the proposed GNN-based solution for problem (\ref{eq:litdiff}). Then, we will explain its training process. The proposed GNN-based algorithm consists of four components: a) input layer, b) hidden layer \uppercase\expandafter{\romannumeral1}, c) hidden layer \uppercase\expandafter{\romannumeral2}, d) hidden layers \uppercase\expandafter{\romannumeral3}-\uppercase\expandafter{\romannumeral5}, and e) output layer, which are specified as follows:

\begin{itemize} 
    \item \emph{Agent}: The agent is a central controller that can obtain the geographic locations of all vehicles. In each time slot, the central controller implements the proposed GNN-based algorithm to determine the service mode and target vehicle association for each SPV. Therefore, the controller actually executes the neural network $|\cal{K}|$ times so as to determine the service mode and target vehicle association for the $|\cal{K}|$ vehicles. 
    
    \begin{figure}[t]
    \centering
    \vspace{-0.3cm}
    \includegraphics[width=11cm]{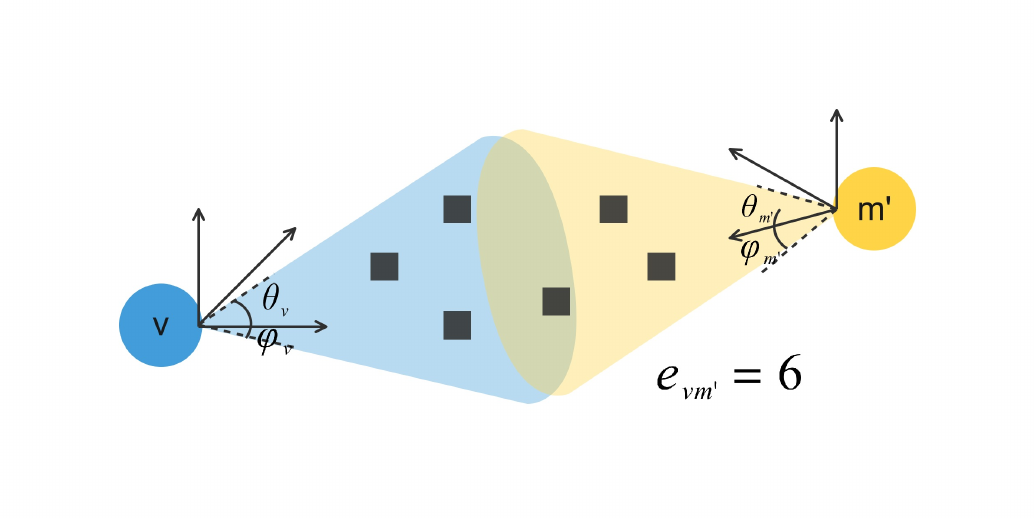}
    \caption{\normalsize An illustration of node feature.}
    \vspace{-0.3cm}
    \label{Beam}
    \end{figure}

   \item \emph{Input Layer}: To determine the service mode of vehicle $k$ and its serviced vehicles, the input of the designed scheme is based on the features of the vehicles that can connect to vehicle $k$. However, since the number of vehicles that can connect to different SPVs are varying, the size of the input matrix may also be different. To enable a neural network to extract graph information for different SPVs that may connect to different number of service target vehicles, we use uniform sampling to calculate the average features of each connected vehicle so as to fix the size of the input. In particular, we assume that the number of vehicles that the proposed algorithm needs to sample for a vehicle $k$ is $s_i$ in sampling iteration $i \in \{1, \cdots, I\}$. Meanwhile, we assume that the set of sampled vehicles that can directly connect to vehicle $k$ as the set of the first hop vehicles, which is represented by ${\cal{L}}^1\left(k\right)$ with $|{\cal{L}}^1(k)|=s_1$ being the number of vehicles in set ${\cal{L}}^1\left(k\right)$. The set of sampled vehicles that can connect to vehicle $k$ via the first hop vehicles as the set of the second hop vehicles represented by ${\cal{L}}^2\left(k\right)$, where  ${\cal{L}}^2\left(k\right)=\{{\cal{L}}^1\left(v^{\prime}\right)|v^{\prime} \in {\cal{L}}^1\left(k\right)\}$. The number of vehicles in set ${\cal{L}}^2\left(k\right)$ is $s_2$. For example, in Fig.~\ref{Sampling}, the total number of sampled vehicles is 5 ( $s_1=2$ and $s_2=3$), ${\cal{L}}^1\left(k\right)=\left\{v_1, v_2\right\}$, and ${\cal{L}}^2\left(k\right)=\{v_3, v_4, v_5\}$. We assume that the subset of the first hop vehicles with a type $R$ edge is ${\cal{L}}_{R}^1\left(k\right)$. For example, in Fig. \ref{Sampling}, ${\cal{L}}_{R_\emph{I}}^1\left(k\right)=\{v_1\}$ and ${\cal{L}}_{R_\emph{SS}}^1\left(k\right)=\{v_2\}$. Given these definitions, we next introduce the input of the proposed GNN-based method. From Fig. \ref{nn}, we see that the input is connected to four fully connected layers and each fully connected layer has different inputs. The inputs to the four fully connected layers are: {\romannumeral1}) $\boldsymbol{h}_k^0=\boldsymbol{f}_k  \in \mathbb{R}^{L \times 1}$, {\romannumeral2}) $\boldsymbol{h}_{R_\emph{SC}}^{1} \in \mathbb{R}^{(L+1)  \times 1}$, {\romannumeral3}) $\boldsymbol{h}_{R_\emph{SS}}^{1} \in \mathbb{R}^{(L+1)  \times 1}$, and {\romannumeral4}) $\boldsymbol{h}_{R_\emph{I}}^{1} \in \mathbb{R}^{(L+1) \times 1}$, where

    \begin{equation}
    \label{input2-1}
    \boldsymbol{h}_{R_\emph{SC}}^{1}=\frac{1}{|{\cal{L}}^1_{R_\emph{SC}}\left(k\right)|}\sum_{v^{\prime} \in {\cal{L}}^1_{R_\emph{SC}}\left(k\right)}{\boldsymbol{h}_{kv^{\prime}}^{0}},
    \end{equation}
    
    \begin{equation}
    \label{input2-2}
    \boldsymbol{h}_{R_\emph{SS}}^{1}=\frac{1}{|{\cal{L}}^1_{R_\emph{SS}}\left(k\right)|}\sum_{v^{\prime} \in {\cal{L}}^1_{R_\emph{SS}}\left(k\right)}{\boldsymbol{h}_{kv^{\prime}}^{0}},
    \end{equation}
    
    \begin{equation}
    \label{input2-3}
    \boldsymbol{h}_{R_\emph{I}}^{1}=\frac{1}{|{\cal{L}}^1_{R_\emph{I}}\left(k\right)|}\sum_{v^{\prime} \in {\cal{L}}^1_{R_\emph{I}}\left(k\right)}{\boldsymbol{h}_{kv^{\prime}}^{0}},
    \end{equation}
    with ${\boldsymbol{h}_{kv^{\prime}}^{0}}=\left[\boldsymbol{h}_{v^{\prime}}^0 \| g_{kv^{\prime}}\right]$, $\boldsymbol{h}_{kv^{\prime}}^{0} \in \mathbb{R}^{(L+1) \times 1}$, $\cdot \| \cdot$ representing the vector concatenation operation, $|{\cal{L}}^1_{R_\emph{SC}} \left(k \right)|$, $|{\cal{L}}^1_{R_\emph{SS}} \left(k \right)|$, and $|{\cal{L}}^1_{R_\emph{I}} \left(k\right)|$ being the number of vehicles in set ${\cal{L}}^1_{R_\emph{SC}} \left(k \right)$, ${\cal{L}}^1_{R_\emph{SS}} \left(k \right)$, and ${\cal{L}}^1_{R_\emph{I}} \left(k \right)$, respectively.

    \begin{figure}[t]
    \centering
    \vspace{-0.3cm}
    \includegraphics[width=11cm]{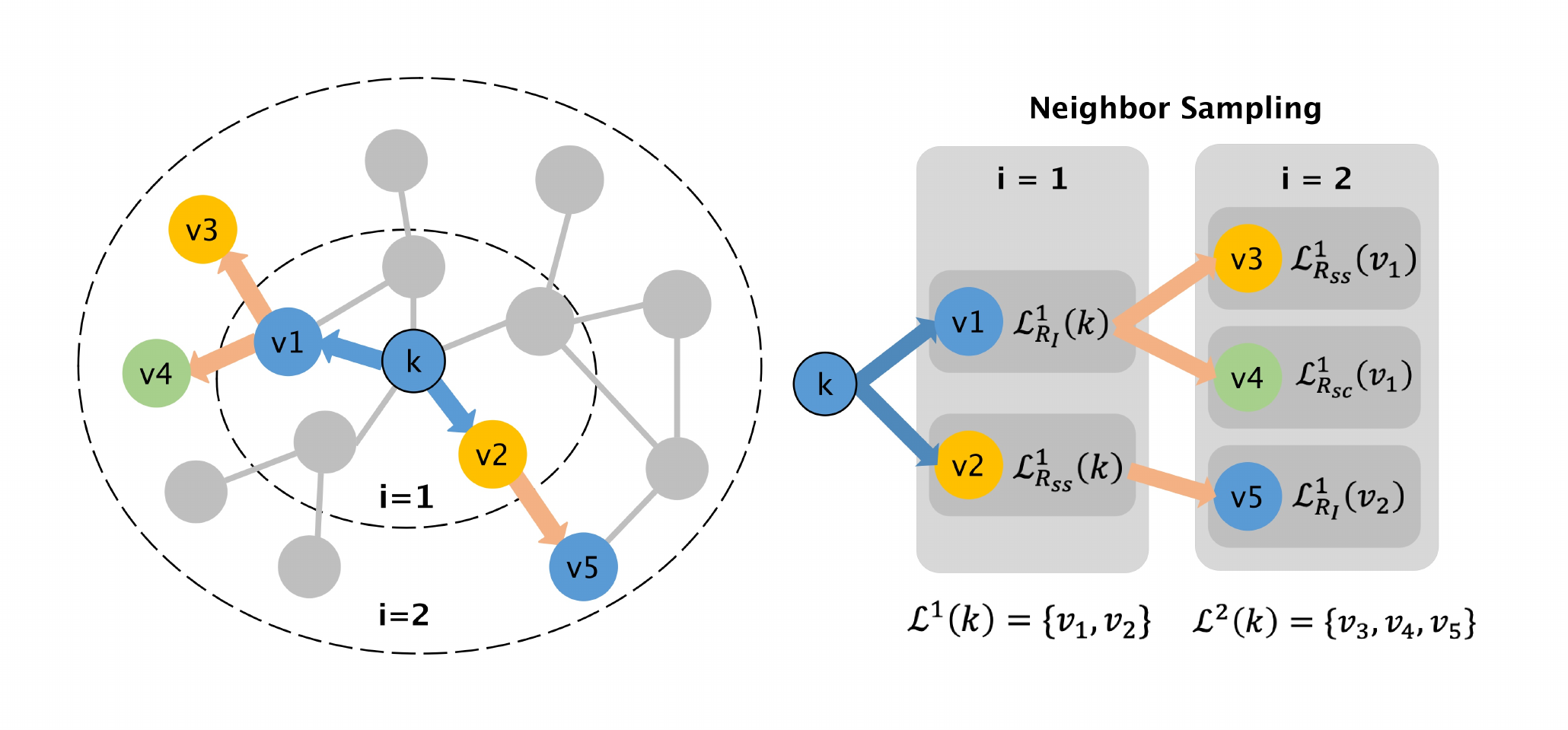}
    \caption{\normalsize Aggregate feature information from neighbors.}
    \vspace{-0.1cm}
    \label{Sampling}
    \end{figure}
    
    \item \emph{Hidden Layer \uppercase\expandafter{\romannumeral1}}: This layer consists of four fully-connected layers and it is used to extract the graph information of the first hop vehicles of each vehicle $k$. The output of this layer is 
     \begin{equation}
     \label{con1}
      {\boldsymbol{h}_{k}^{1}} = \sigma\left( \left[{\boldsymbol{w}_1}{\boldsymbol{h}_{k}^{0}} \|
      {\boldsymbol{w}_2} {\boldsymbol{h}^{1}_{{R_\emph{SC}}}} \|
      {\bm{w}_3} {\boldsymbol{h}^{1}_{{R_\emph{SS}}}}\|
      {\boldsymbol{w}_4} {\boldsymbol{h}^{1}_{{R_\emph{I}}}}
      \right]\right), 
     \end{equation}
    where $\sigma\left(\cdot\right)$ is the rectified linear unit function (ReLU), $\boldsymbol{w}_1 \in \mathbb{R}^{\left(\lambda_{0}/4 \right) \times L}$, $\boldsymbol{w}_2 \in \mathbb{R}^{\left(\lambda_{0}/4 \right) \times \left(L+1 \right)}$, $\boldsymbol{w}_3 \in \mathbb{R}^{\left(\lambda_{0}/4 \right) \times \left(L+1 \right)}$, and $\boldsymbol{w}_4 \in \mathbb{R}^{\left(\lambda_{0}/4 \right) \times \left(L+1 \right)}$ are the weights of the four fully connected layers, $\lambda_{0}$ is the dimension of the graph information vector, $\boldsymbol{w}_1$ is the weight matrix of the current vehicle, and $\boldsymbol{w}_2$, $\boldsymbol{w}_3$, and $\boldsymbol{w}_4$ are the weight matrices for the vehicles with the type ${R_\emph{SC}}$, ${R_\emph{SS}}$, and ${R_\emph{I}}$ edge, respectively. To support heterogeneous nodes and edges, we set separate neighbourhood weight  matrices $\boldsymbol{w}_2$, $\boldsymbol{w}_3$, and $\boldsymbol{w}_4$ for each type of vehicles, respectively. From (\ref{input2-1}) to (\ref{con1}), we extract only the graph information of vehicle $v$. However, we need the graph information of all the sampled first hop vehicles to optimize service mode selection and vehicle connection. Therefore, we need to execute (\ref{input2-1}) to (\ref{con1}) for each sampled vehicle (i.e., for $s_1$ times). After that, we obtain  $\boldsymbol{h}_{v^{\prime}}^{1} \in \mathbb{R}^{\lambda_{0} \times 1}$, $ \forall v^{\prime} \in {\cal{L}}^1 \left(k\right)$ for each sampled vehicle $v^{\prime}$.
  
   \item \emph{Hidden Layer  \uppercase\expandafter{\romannumeral2}}: This layer consists of four fully-connected layers and it is used to extract the graph information of the second hop vehicles of vehicle $k$. From Fig. \ref{nn}, we can see that the input to each fully connected layer in hidden layer \uppercase\expandafter{\romannumeral2} is different. The inputs to the four fully connected layers are: {\romannumeral1}) $\boldsymbol{h}_k^1  \in \mathbb{R}^{\lambda_{0} \times 1}$, {\romannumeral2}) $\boldsymbol{h}_{R_\emph{SC}}^{2} \in \mathbb{R}^{\left(\lambda_{0}+1 \right)  \times 1}$, {\romannumeral3}) $\boldsymbol{h}_{R_\emph{SS}}^{2} \in \mathbb{R}^{\left(\lambda_{0}+1 \right)  \times 1}$, and {\romannumeral4}) $\boldsymbol{h}_{R_\emph{I}}^{2} \in \mathbb{R}^{\left(\lambda_{0}+1\right) \times 1}$, where
       
    \begin{equation}
    \label{input3-1}
    \boldsymbol{h}_{R_\emph{SC}}^{2}=\frac{1}{|{\cal{L}}^1_{R_\emph{SC}}\left(k\right)|}\sum_{v^{\prime} \in {\cal{L}}^1_{R_\emph{SC}} \left(k\right)}{\boldsymbol{h}_{kv^{\prime}}^{1}},
    \end{equation}
    
     \begin{equation}
    \label{input3-2}
    \boldsymbol{h}_{R_\emph{SS}}^{2}=\frac{1}{|{\cal{L}}^1_{R_\emph{SS}}\left(k\right)|}\sum_{v^{\prime} \in {\cal{L}}^1_{R_\emph{SS}}\left(k\right)}{\boldsymbol{h}_{kv^{\prime}}^{1}},
    \end{equation}
    
     \begin{equation}
    \label{input3-3}
    \boldsymbol{h}_{R_\emph{I}}^{2}=\frac{1}{|{\cal{L}}^1_{R_\emph{I}}\left(k\right)|}\sum_{v^{\prime} \in {\cal{L}}^1_{R_\emph{I}}\left(k\right)}{\boldsymbol{h}_{kv^{\prime}}^{1}},
    \end{equation}
    with ${\boldsymbol{h}_{kv^{\prime}}^{1}}=\left[\boldsymbol{h}_{v^{\prime}}^1 \| g_{kv^{\prime}}\right]$ and $\boldsymbol{h}_{kv^{\prime}}^{1} \in \mathbb{R}^{\left(\lambda_{0}+1\right) \times 1}$. The output of this layer is
    \begin{equation}
    \label{con2}
      {\boldsymbol{h}_{k}^{2}} = \sigma\left( \left[{\boldsymbol{w}_5}{\boldsymbol{h}_{k}^{1}} \|
      {\boldsymbol{w}_6} {\boldsymbol{h}^{2}_{{R_\emph{SC}}}} \|
      {\bm{w}_7} {\boldsymbol{h}^{2}_{{R_\emph{SS}}}} \|
      {\boldsymbol{w}_8} {\boldsymbol{h}^{2}_{{R_\emph{I}}}}
      \right]\right), 
     \end{equation}
    where $\boldsymbol{h}_k^2 \in   \mathbb{R}^{\lambda_{0} \times 1}$, $\boldsymbol{w}_5 \in \mathbb{R}^{\left(\lambda_{0}/4 \right) \times \lambda_{0}}$, $\boldsymbol{w}_6 \in \mathbb{R}^{\left(\lambda_{0}/4 \right) \times \left(\lambda_{0}+1\right)}$, $\boldsymbol{w}_7 \in \mathbb{R}^{\left(\lambda_{0}/4\right) \times \left(\lambda_{0}+1\right)}$, and $\boldsymbol{w}_8 \in \mathbb{R}^{\left(\lambda_{0}/4\right) \times \left(\lambda_{0}+1\right)}$ are the weights of the four fully connected layers, respectively. $\boldsymbol{w}_5$ is the weight matrix of vehicle $k$ and the others are the weight matrices for the three types of second hop vehicles, i.e., $\boldsymbol{w}_6$ is the weight matrix for the vehicles with a type ${R_\emph{SC}}$ edge, $\boldsymbol{w}_7$ is the weight matrix for the vehicles with a type ${R_\emph{SS}}$ edge, and $\boldsymbol{w}_8$ is the weight matrix for the vehicles with a type ${R_\emph{I}}$ edge. Compared to the aggregate function in \cite{StellarGraph} and \cite{ Hamilton2017Inductive} that considers only node features, we consider both node features and edge weights in both hidden layers \uppercase\expandafter{\romannumeral1} and \uppercase\expandafter{\romannumeral2}. Here, the output ${\boldsymbol{h}_{k}^{2}}$ can be considered as the graph information of vehicle $k$, since it includes the graph information of the sampled first hop and second hop vehicles.
    
    \item \emph{Hidden Layers \uppercase\expandafter{\romannumeral3}-\uppercase\expandafter{\romannumeral5}}: Three fully-connected layers are used to find the relationship between the graph information vector $\boldsymbol{h}_k^2$ and the probability distribution of vehicle $k$ servicing each target vehicle in the corresponding operating mode. 
   
      
   
     \item \emph{Output}: The output of the network, $\bm{y}_{k}=\left[y_{k}^{1}, \cdots ,y_{k}^{L+1}\right]$, is the probability distribution of vehicle $k$ servicing $L+1$ target vehicles in the corresponding operating mode. Here, $L+1$ is the total number of classification classes, including the case that vehicle $k$ is not connected to any target vehicles. 
\end{itemize}

\subsection{Training the Proposed GNN-based Model}
Given the components defined in the previous section, we next introduce the entire procedure of training the proposed GNN-based method. We use binary cross entropy (BCE) as the loss function to minimize the difference between the predicted multi-label classification result and the actual multi-label classification result, which is given by:
    \begin{equation}
    \label{BCEloss}
    J\left(\boldsymbol{w},\boldsymbol{p},\boldsymbol{b}\right)= \sum_{l=1}^{L+1} -z^{l}_{k} \log \delta \left({y}^{l}_{k}\right)-\left(1-z^{l}_{k}\right) \log \left(1-\delta \left({y}^{l}_{k}\right)\right),
    \end{equation}
where $\delta\left(\cdot\right)$ is the sigmoid function; $z^{l}_{k}$ is the label of  vehicle $k$ for class $l$, which is generated by exhaustive searching; $\boldsymbol{w}$ is the weight matrix of hidden layer \uppercase\expandafter{\romannumeral1}; and 
$\boldsymbol{p}$ and $\boldsymbol{b}$ are the weight matrix and bias of hidden layer \uppercase\expandafter{\romannumeral3}-\uppercase\expandafter{\romannumeral5}, respectively. To minimize the training loss (\ref{loss}), we optimize $\boldsymbol{w}$, $\boldsymbol{p}$, and $\boldsymbol{b}$ using the back-propagation algorithm with the mini-batch stochastic gradient descent (SGD) approach \cite{Hamilton2017Inductive}. The parameters of each fully-connected layer $j$ in the GNN, i.e., $\boldsymbol{w}_j$, $\boldsymbol{p}_{j^{\prime}}$, and $\boldsymbol{b}_{j^{\prime}}$, are randomly initialized by a uniform distribution and updated by the central controller in each training iteration $t \in \{1, \cdots, T\}$ of the  mini-batch SGD approach, where $T$ is the number of total training iterations. Specifically, the standard update policy of  mini-batch SGD is given by:

\begin{equation}
\label{update1-1}
\boldsymbol{w}_j^{t+1}=\boldsymbol{w}_j^{t}-\eta\frac{1}{|\mathcal{B}|}\sum_{k\in \mathcal{B}_t} \frac{\partial J\left(\boldsymbol{w},\boldsymbol{p},\boldsymbol{b}\right)}{\partial\boldsymbol{w}_j},
\end{equation}

\begin{equation}
\label{update1-2}
\boldsymbol{p}^{t+1}_{j^{\prime}}=\boldsymbol{p}^{t}_{j^{\prime}}-\eta\frac{1}{|\mathcal{B}|}\sum_{k\in \mathcal{B}_t} \frac{\partial J\left(\boldsymbol{w},\boldsymbol{p},\boldsymbol{b}\right)}{\partial\boldsymbol{p}_{j^{\prime}}},
\end{equation}

\begin{equation}
\label{update1-3}
\boldsymbol{b}^{t+1}_{j^{\prime}}=\boldsymbol{b}^{t}_{j^{\prime}}-\eta\frac{1}{|\mathcal{B}|}\sum_{k\in \mathcal{B}_t} \frac{\partial J\left(\boldsymbol{w},\boldsymbol{p},\boldsymbol{b}\right)}{\partial\boldsymbol{b}_{j^{\prime}}},
\end{equation}
where ${|\mathcal{B}|}$ is the size of mini-batches, $\mathcal{B}_t$ is a mini-batch of training samples used in iteration $t$, $\eta $ is the learning rate, $j \in \{1,2,3,4,5,6,7,8\}$, and ${j^{\prime} \in \{1,2,3\}}$. $\boldsymbol{w}_j, \forall j \in \{1,2,3,4\}$ is the weight matrix of fully-connected layer $j$ in hidden layer \uppercase\expandafter{\romannumeral1}. $\boldsymbol{w}_j, \forall j \in \{5,6,7,8\}$ is the weight matrix of fully-connected layer $j$ in hidden layer \uppercase\expandafter{\romannumeral2}. $\boldsymbol{p}_{j^{\prime}}, \forall j^{\prime} \in \{1,2,3\}$ and $b_{j^{\prime}}, \forall {j^{\prime}} \in \{1,2,3\}$ are the weight matrix and bias of fully-connected layer $j^{\prime} $ in hidden layer \uppercase\expandafter{\romannumeral3}-\uppercase\expandafter{\romannumeral5}. The entire training process of the proposed GNN-based algorithm is summarized in \textbf{Algorithm 1}.


\begin{algorithm}[t]
\caption{GNN-based Method for the Joint Service Mode Selection and Target Vehicle Association Problem}
\begin{algorithmic}[1]
\STATE \textbf{Input:} Vehicle features $\{\boldsymbol{f}_v, \forall v \in {\cal{V}} \}$, edge weights $\{g_{v{v^{\prime}}}, \forall v^{\prime} \in {\cal{V}}\setminus\{v\}\}$, and sampling size $s_1$ and $s_2$;

\STATE \textbf{Initialize:} $\boldsymbol{w}_j$, $\boldsymbol{p}_{j^{\prime}}$, and $b_{j^{\prime}}$; 

\STATE $\boldsymbol{h}_v^0 \leftarrow \boldsymbol{f}_v$, ${\boldsymbol{h}_{vv^{\prime}}^{0}} \leftarrow \left[\boldsymbol{h}_{v^{\prime}}^0 \| g_{vv^{\prime}}\right]$, $\forall v \in {\cal{V}},\forall v^{\prime} \in {\cal{V}}\setminus\{v\}$;

\FOR {$k = 1 \to K$}

\STATE Sample the first hop vehicles ${\cal{L}}^1 \left(k\right)$ and second 
hop vehicles ${{\cal{L}}^2\left(k\right)}$ for vehicle $k$;

\STATE  Extract the graph information $\boldsymbol{h}_k^1$ of vehicle $k$ based on (\ref{input2-1})-(\ref{con1});

\FOR {$v^{\prime} \in {\cal{L}}^1 \left(k\right)$}

\STATE Extract the graph information ${\bm{h}_{v^{\prime}}^1}$ of vehicle $v^{\prime}$ based on (\ref{input2-1})-(\ref{con1});

\ENDFOR
\STATE  ${\boldsymbol{h}_{kv^{\prime}}^{1}} \leftarrow [\boldsymbol{h}_{v^{\prime}}^1 \| g_{kv^{\prime}}]$, $\forall v^{\prime} \in {\cal{L}}^1\left(k\right)$;

\STATE  Aggregate the neighboring feature vectors of vehicle $k$, $\boldsymbol{h}_{R_\emph{SC}}^{2}$, $\boldsymbol{h}_{R_\emph{SS}}^{2}$, and        $\boldsymbol{h}_{R_\emph{I}}^{2}$, based on (\ref{input3-1})-(\ref{input3-3});

\STATE Concatenate the vehicle’s current representation, $\boldsymbol{h}_k^1$, with the aggregated neighborhood vector based on (\ref{con2});

\STATE Obtain the graph information vector  ${\bm{h}_{k}^2}$ for vehicle $k$;

\STATE Use ${\bm{h}_{k}^2}$ as input to predict the probability distribution ${\boldsymbol{y}_k}$ of vehicle $k$;

\STATE Calculate loss $J\left(\boldsymbol{w},\boldsymbol{p},\boldsymbol{b}\right)$ based on (\ref{BCEloss});

\STATE Update the weight matrices and bias using (\ref{update1-1})-(\ref{update1-3});

\ENDFOR

\STATE \textbf{Output:} The probability distribution $\boldsymbol{y}_k$ for each vehicle $k \in {\cal{K}}$.
\end{algorithmic}
\label{algorithm_2}
\end{algorithm}

\subsection{Solution for the Formulated Problem}

Once the probability distribution $\boldsymbol{y}_k$ of each SPV $k$ is obtained, the service mode selection and target vehicle association can be determined. In particular, given $\boldsymbol{y}_k$, the objective function in (\ref{eq:litdiff}) can be approximated by   
\begin{equation}
\label{association_pro_add} 
E(\bm{\alpha},\bm{\beta})=\sum_{m\in \mathcal{M}} \sum_{k\in \mathcal{K}}
 \alpha_{k m} y_{k}^{m}+ \sum_{n\in \mathcal{N}} \sum_{k\in \mathcal{K}} \beta _{kn} y_{k}^{{|\cal{M}|}+n},
\end{equation}
where the first term $\sum_{m\in \mathcal{M}} \sum_{k\in \mathcal{K}} \alpha_{k m} y_{k}^{m}$ is the sum of the probabilities of establishing all the communication links, and the second term $\sum_{n\in \mathcal{N}} \sum_{k\in \mathcal{K}} \beta _{kn} y_{k}^{{|\cal{M}|}+n}$ represents the sum of the probabilities of establishing all the sensing links. Here, a small gap may exist between the original objective function in (\ref{eq:litdiff}) and the approximated objective function in (\ref{association_pro_add}). However, this approximation can significantly simplify the solution procedure for determining service mode selection and target vehicle association. We will use simulation results in Section \ref{section:S} to verify the accuracy of this approximation process. Given (\ref{association_pro_add}), the problem in (\ref{eq:litdiff}) can be rewritten as
 \begin{subequations}\label{eq:obj_pro}
\begin{align}
	&\mathop{\mbox{max}}_{\bm{\alpha},\bm{\beta}}~E(\bm{\alpha},\bm{\beta}) \tag{\ref{eq:obj_pro}}\\
	\mbox{s.t.} &\sum_{k\in  \mathcal{K}}\alpha_{km}=1,\alpha_{km} \in \{0,1\},\forall m \in \mathcal{M},\label{eq:25a} \\
	&\sum_{k\in  \mathcal{K}}\beta_{kn}=1,\beta _{kn} \in \{0,1\},\forall n \in \mathcal{N},\label{eq:25b} \\
	&\sum_{m\in  \mathcal{M}} \alpha_{km}\geq 0, \sum_{n\in  \mathcal{N}}\beta_{kn}\geq 0, \forall k \in \mathcal{K},\label{eq:25c}\\
	&\alpha_{km}\beta _{kn}=0, \forall k \in \mathcal{K},\forall m \in \mathcal{M},\forall n \in \mathcal{N},\label{eq:25d}\\
	&\sum_{k\in  \mathcal{K}} \mathrm{\gamma}_{kn}^{\textrm{S}}(\bm{\alpha},\bm{\beta}) \geq \gamma_{\min },\forall n \in \mathcal{N}. \label{eq:25e}
\end{align}
\end{subequations}
 In (\ref{eq:obj_pro}), constraints (\ref{eq:25d}) and (\ref{eq:25e}) are non-linear. Therefore, we need to rewrite and linearize these two constraints. For constraint (\ref{eq:25d}), we can rewrite it as $\alpha_{km}+\beta _{kn} \leq 1$, which guarantees that a service provider
vehicle can only operate in either the communication mode or the sensing mode. For constraint (\ref{eq:25e}), since a sensing target vehicle $n$ can only be connected with one service provider
vehicle $k$, i.e., $\sum_{k\in  \mathcal{K}} \mathrm{\gamma}_{kn}^{\textrm{S}}(\bm{\alpha},\bm{\beta})=\mathrm{\gamma}_{kn}^{\textrm{S}}(\bm{\alpha},\bm{\beta})$ when $\beta_{kn}=1$, we have $\mathrm{\gamma}_{kn}^{\textrm{S}}(\bm{\alpha},\bm{\beta}) \geq \beta _{kn}\gamma_{\min }$.

Given the rewritten constraints (\ref{eq:25d}) and (\ref{eq:25e}), the optimization problem in (\ref{association_pro_add}) is now given by
\begin{subequations}\label{eq:obj_pro2}
\begin{align}
	\mathop{\mbox{max}}_{\bm{\alpha},\bm{\beta}} & E(\bm{\alpha},\bm{\beta}) \tag{\ref{eq:obj_pro2}}\\
	\mbox{s.t.} &(\ref{eq:25a}) - (\ref{eq:25c}), \\
	&\alpha_{km}+\beta _{kn} \leq 1, \forall k \in \mathcal{K},\forall m \in \mathcal{M},\forall n \in \mathcal{N},\label{eq:26d}\\
	&\mathrm{\gamma}_{kn}^{\textrm{S}}(\bm{\alpha},\bm{\beta}) \geq \beta _{kn}\gamma_{\min },\forall k \in \mathcal{K},\forall n \in \mathcal{N}.\label{eq:26e}
\end{align}
\end{subequations}
Problem (\ref{eq:obj_pro2}) is a quadratically constrained programming (QCP) problem. Thus, it can be solved by using a convex optimization tool, such as Gurobi \cite{Gurobi}.

\subsection{Implementation of the Proposed GNN-based Algorithm}
Next, we analyze the implementation of the GNN-based algorithm. To implement the GNN-based algorithm for finding the optimal service mode selection and target vehicle association matrices $\bm{\alpha}$ and $\bm{\beta}$, the central controller must first obtain the vehicle topology. The vehicle topology is constructed based on the vehicles’ periodically reported GPS data. Then, we need to transform the vehicle topology into the heterogeneous graph representation. To establish the heterogeneous graph representation, the central controller must know 1) the path loss (i.e., the spreading loss and the absorption loss) between each two vehicles $v ,v^{\prime} \in {\cal{V}}$ to obtain the edge feature ${g}_{vv^{\prime}}$, and 2) the antenna direction between each vehicle $v \in {\cal{V}}$ and each service target vehicle $m^{\prime} \in {\cal{M}^{\prime}}$  to obtain the node feature $\boldsymbol{f}_{v}$. The central controller can use channel estimation methods to learn the path loss and the antenna direction of each pair of vehicles \cite{Huang2020Fair}. Based on the heterogeneous graph representation, a GNN is used to determine the probability distribution $\bm{y}_k$ for each SPV. Given $\bm{y}_k$, the convex optimization tool can be used to find the optimal service mode selection and target vehicle association matrices $\bm{\alpha}$ and $\bm{\beta}$. Since the optimization function in (\ref{eq:obj_pro2}) is convex, it will finally find the optimal $\bm{\alpha}$ and $\bm{\beta}$.

The proposed GNN-based algorithm includes an offline training phase and an online decision making phase. A well trained GNN model is transferred from the offline phase to the online phase. In the offline phase, the model is trained during the idle time of the central controller leveraging the historical geographic locations and network topological information. When the GNN model is well trained, a series of reasonable weights that can accurately map an input to an output are obtained. During the online decision making phase, the trained weights can be directly used for generating the probability distribution of each SPV without updating the weights of GNNs. 

\subsection{Complexity Analysis}
Next, we analyze the computational complexity of the proposed GNN-based scheme for service mode selection and target vehicle association optimization. The complexity of the proposed scheme is analyzed from two parts: 1) complexity for training GNN-based scheme and 2) complexity for inference. The training process of the proposed GNN model is conducted once in the offline training phase, while the inference process of determining the service mode of each service provider vehicle and target vehicle association is conducted for each vehicle topology.

 \subsubsection{Complexity for Training GNN-based Scheme}
 The complexity for training GNN-based scheme lies in computing the graph information vector and performing multi-label classification. The computational complexity for graph information vector computation depends on the size of neighbor sampling $s_i$, the number of sampling iteration $I$, the dimension of graph information vector $\lambda_{0}$, and the number of SPVs ${|\cal{K}|}$. Hence, the computational complexity for graph information vector computation is given by
${\cal{O}} \left(\lambda_{0} {|\cal{K}|} {\prod_{i=1}^{I} s_i}\right)$. According to \cite{He2015Convolutional}, the complexity of training a neural network depends on its width, depth, and number of parameters. The complexity of training hidden layer \uppercase\expandafter{\romannumeral3}-\uppercase\expandafter{\romannumeral5} to perform multi-label classification is ${\cal{O}} \left(\prod_{i=2}^{J} H_j\right)$, where $H_j$ is the number of the neurons in layer $j$ and $J$ is the number of layers. Therefore, the computational complexity for training GNN-based scheme is given by
\begin{equation}
\label{complexity}
{\cal{O}} \left(\left( \lambda_{0} {|\cal{K}|} {\prod_{i=1}^{I} s_i} \right)\prod_{i=2}^{J} H_j  \right).
\end{equation}
 
 \subsubsection{Complexity for Inference }
 The complexity for inference lies in determining the service mode selection and target vehicle association strategy based on the multi-label classification results. To solve problem (\ref{eq:obj_pro2}), the computational complexity for determining the service mode selection and target vehicle association matrices is ${\cal{O}} \left(C{|\cal{K}|}{|\cal{M}|}{|\cal{N}|}\right) \approx {\cal{O}} \left({|\cal{K}|}{|\cal{M}|}{|\cal{N}|}\right)$, where $C$ denotes the number of multiplications per search step. It can be seen that given the structure of the proposed GNN-based model (i.e., $s_i$, $H_j$, $\lambda_{0}$, and $I$), the complexity of inference depends on the number of SPVs, the number of communication target vehicles, and the number of sensing target vehicles. 

\section{Simulation Results}
\label{section:S}
\begin{figure}[h]
\centering
\includegraphics[width=11cm]{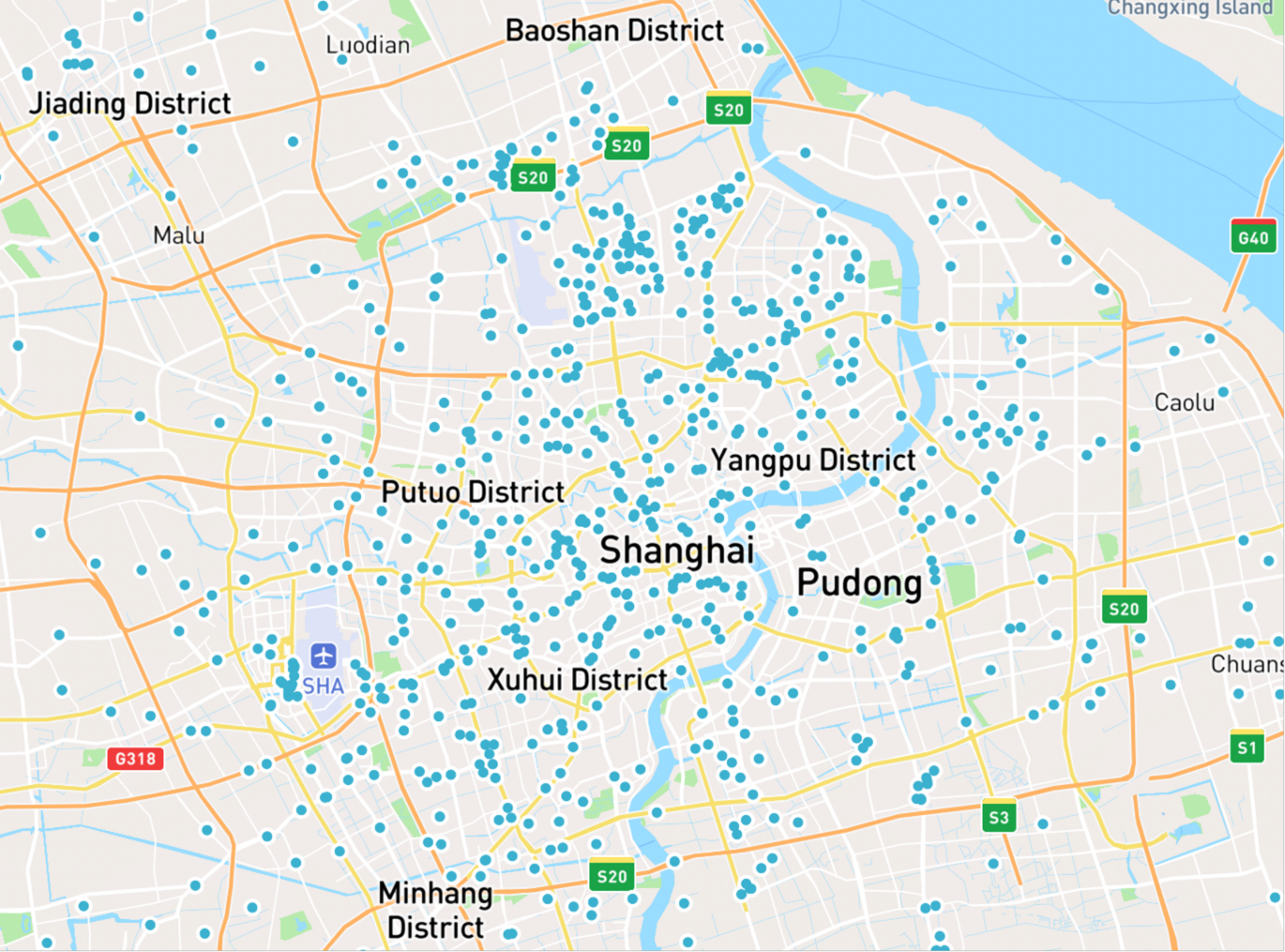}
\caption{\normalsize Visualization of the GPS data.}
\label{Road}
\end{figure}

In this section, we perform extensive simulations to evaluate the performance of our proposed scheme in a specific region of 100 m × 100 m. The other detailed parameters are listed in Table \uppercase\expandafter{\romannumeral2}\cite{Shafie2021Coverage,Wu2021Interference}. The GPS dataset used to generate vehicle topologies is obtained from Shanghai Traffic Department, which consists of ID, timestamp, latitude and longitude information of 18,900 vehicles \cite{Zhao2017Prediction}. In Fig. \ref{Road}, we show a visualization of the GPS data to illustrate the distribution of vehicles. For comparison purposes, we consider three baselines. Baseline a is an exhaustive search algorithm, which can be considered as the optimal solution for problem (\ref{eq:litdiff}). Baseline b is based on homogeneous graph. For a fair comparison, Baseline b uses the same neural network architecture as the proposed method, but a different graph information extraction method given in \cite{He2020Resource}, where different types of nodes and edges are not distinguished. Baseline c directly uses the geographic location information to optimize service mode selection and target vehicle association scheme, without using GNNs to extract the graph information vectors. 


\begin{table}
\caption[table]{{\normalsize System Parameters }}
\centering
\begin{tabular}{|c|c|c|c|} 
\hline
\!\textbf{Parameters}\! \!\!& \textbf{Value} &\! \textbf{Parameters} \!& \textbf{Value} \\
\hline
$c$ & $3 \times 10^{8}$ m/s & $f$ & 1.05 THz 
\\
\hline
$P$ & 40 dBm & $B$ & 5 GHz \\
\hline
$N_{0}$  & -77 dBm  & $\phi(f)$ & 0.07512 m$^{-1}$ \\
\hline
$\varepsilon$ & 0.1 & $\theta,\varphi$ & $10^\circ$, $10^\circ$ \\
\hline
$\sigma$ & 1 & $\gamma_{\min}$ & 3 dB \\
\hline
$\lambda_0$ & 64  & $\eta$ & 0.7\\
\hline 
$s_1$ & 10 & $s_2$ & 10 \\
\hline
$I$ & 2 & $T$ & 20,000  \\
\hline 
\multicolumn{3}{|c|}{Number of training vehicle topologies} & 1,500\\
\hline 
\multicolumn{3}{|c|}{Number of testing vehicle topologies} & 1,000 \\
\hline 
\multicolumn{3}{|c|}{Number of validating vehicle topologies} & 1,000\\
\hline 
\multicolumn{3}{|c|}{Size of hidden layer \uppercase\expandafter{\romannumeral3}-\uppercase\expandafter{\romannumeral5}} & 32, 64, 64 \\
\hline 
\multicolumn{3}{|c|}{Size of mini-batches ${|\mathcal{B}|}$} & 64\\
\hline 
\end{tabular}
\end{table}

\begin{figure}[t]
\centering
\includegraphics[width=12cm]{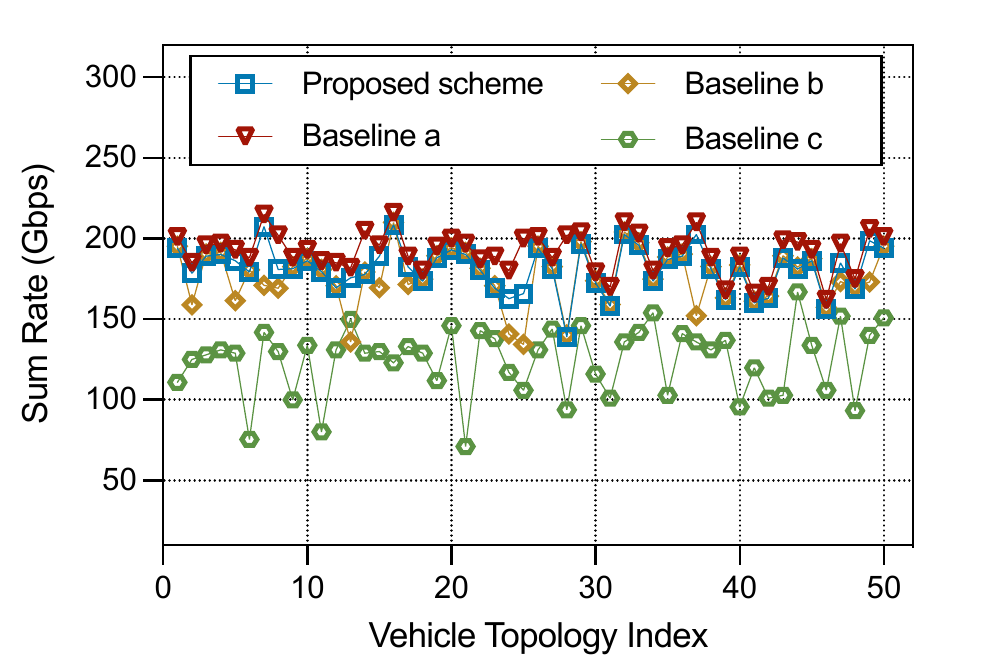}
\caption{\normalsize The sum rate as the vehicle topology varies (${|\mathcal{K}|}=5$, ${|\mathcal{M}|}=2$, and ${|\mathcal{N}|}=2$).}
\label{SumRate}
\end{figure}
Fig. \ref{SumRate} shows how the sum of data rates of all communication target vehicles change as the vehicle topology varies. 
From Fig. \ref{SumRate}, we see that the proposed scheme improves the sum rate by up to 3.16\% and 31.86\% compared to baselines b and c. The 3.16\% gain stems from the fact that the proposed scheme uses a heterogeneous graph to represent a vehicle topology, hence, the impact of different types of vehicles on service mode selection and target vehicle association is considered. The 31.86\% gain stems from the fact that the proposed scheme uses a GNN to extract a graph information vector for each SPV. In Fig. \ref{SumRate}, we can also observe that, there is only a small performance gap between the proposed scheme and baseline a, which verifies the high approximation accuracy of (\ref{eq:obj_pro2}).

\begin{figure}[t]
\centering
\includegraphics[width=12cm]{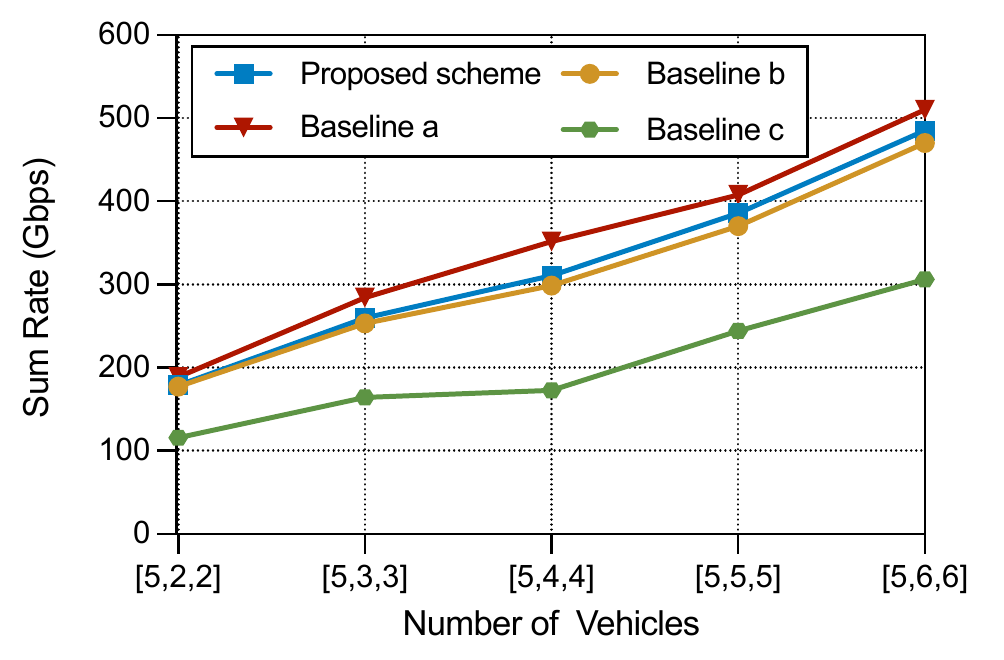}
\caption{\normalsize The sum rate as the number of target vehicles varies (${|\mathcal{K}|}=5$, $|\mathcal{M}|$ and $|\mathcal{N}|$ vary from 2 to 6).}
\label{Rate_Target3}
\end{figure}
Fig.~\ref{Rate_Target3} shows how the sum of data rates of all communication target vehicles changes as the number of communication and sensing target vehicles varies. From this figure, we can see that, as the number of communication and sensing target vehicles increases, the sum of data rates of all communication target vehicles increases since more communication links are established. Fig.~\ref{Rate_Target3} also shows that, compared to baselines b and c, the proposed scheme can achieve up to 2.94\% and 35.45\% gains in terms of the sum rate of all communication target vehicles. The 2.94\% gain stems from the fact that the proposed scheme considers the use of heterogeneous GNNs to extract geographical location information and topological information from different types of vehicles. The 35.45\% gain stems from the fact that the proposed scheme determines the target vehicle association based on the learned graph information vector and hence, it optimizes target vehicle association while considering all vehicle's location, connection, and communication interference. Fig.~\ref{Rate_Target3} also shows that the gap between the proposed scheme and baseline a is less than 7\%. This indicates that the proposed GNN-based scheme enables the trained GNN to adapt to different vehicle topologies with different number of vehicles.

\begin{figure}[t]
\centering
\includegraphics[width=12cm]{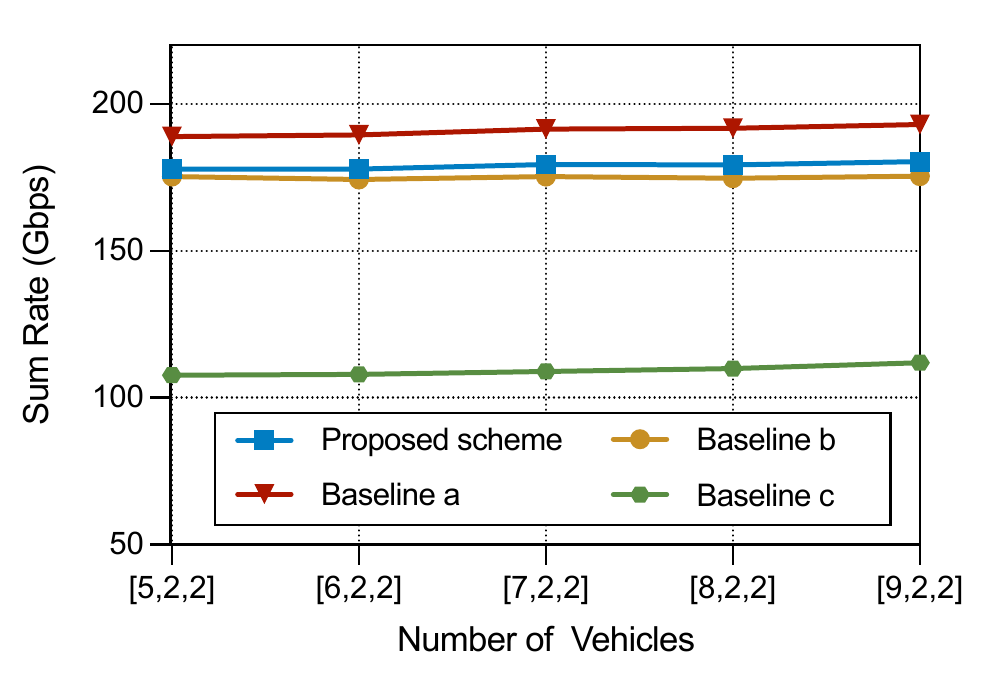}
\caption{\normalsize The sum rate as the number of service vehicles varies (${|\cal{M}|}={|\cal{N}|}=2$ and $|\cal{K}|$ varies from 5 to 9).}
\label{Sum_Service}
\end{figure} 
Fig.~\ref{Sum_Service} shows how the sum of data rates of all communication target vehicles changes as the number of SPVs varies. From this figure, we can see that, as the number of SPVs increases, the sum of data rates of all communication target vehicles increases. This is due to the fact that the increase of the number of SPVs makes more SPVs available for communication target vehicles to select. In consequence, the communication target vehicles are more likely to select the SPVs with an appropriate direction and distance. Fig.~\ref{Sum_Service} also shows that, compared to baseline c, the proposed scheme can achieve up to 36.45\% gain in terms of sum rate. This is because the proposed algorithm considers both geographical location information and topological information. In Fig.~\ref{Sum_Service}, we can also see that the proposed scheme can achieve up to 2.06\% gain in terms of sum rate compared to baseline b. This is due to the fact that the proposed algorithm uses heterogeneou GNN to learn the information of different types of vehicles.

\begin{figure}[t]
\centering
\includegraphics[width=12cm]{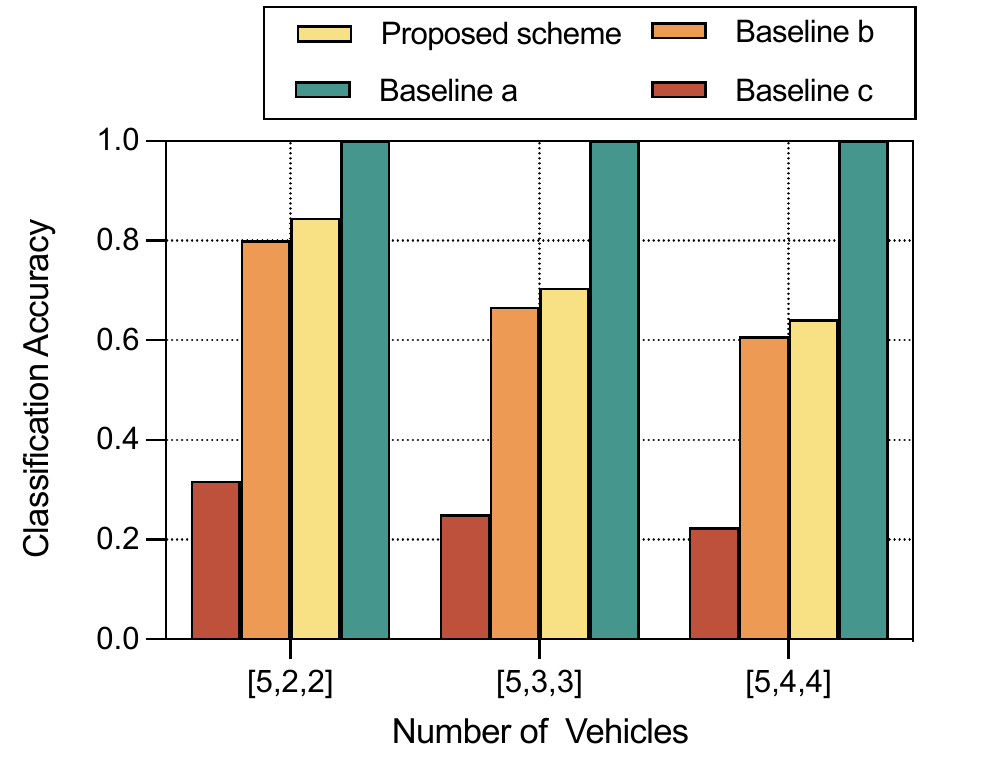}
\caption{\normalsize The classification accuracy as the number of target vehicles varies ($|\mathcal{K}|$=5, $|\mathcal{M}|$ and $|\mathcal{N}|$ varies from 2 to 4).}
\label{Acc_Target}
\end{figure}
Fig.~\ref{Acc_Target} shows how the classification accuracy changes as the number of communication and sensing target vehicles varies. From this figure, we can see that, as the number of communication and sensing target vehicles increases, the classification accuracy resulting from all considered algorithms decreases. This is due to the fact that each SPV may now serve more target vehicles simultaneously. As the number of target vehicles increases, the interference among different vehicles increases and the connections among different vehicles become more complicated. Fig.~\ref{Acc_Target} also shows that, compared to baseline c, the proposed scheme can achieve up to 46.77\% gain in terms of classification accuracy. The reason is that the proposed scheme use GNNs to extract the graph information vectors, which can capture the location, connection, and interference information between vehicles. In Fig.~\ref{Acc_Target}, we can also see that the proposed scheme can achieve up to 4.05\% gain in terms of classification accuracy compared to baseline b. This is due to the fact that the proposed scheme uses separate weight matrices for each type of vehicle, hence, the graph information vectors of different types of vehicles can be better represented. 

\begin{figure}[t]
\centering
\includegraphics[width=12cm]{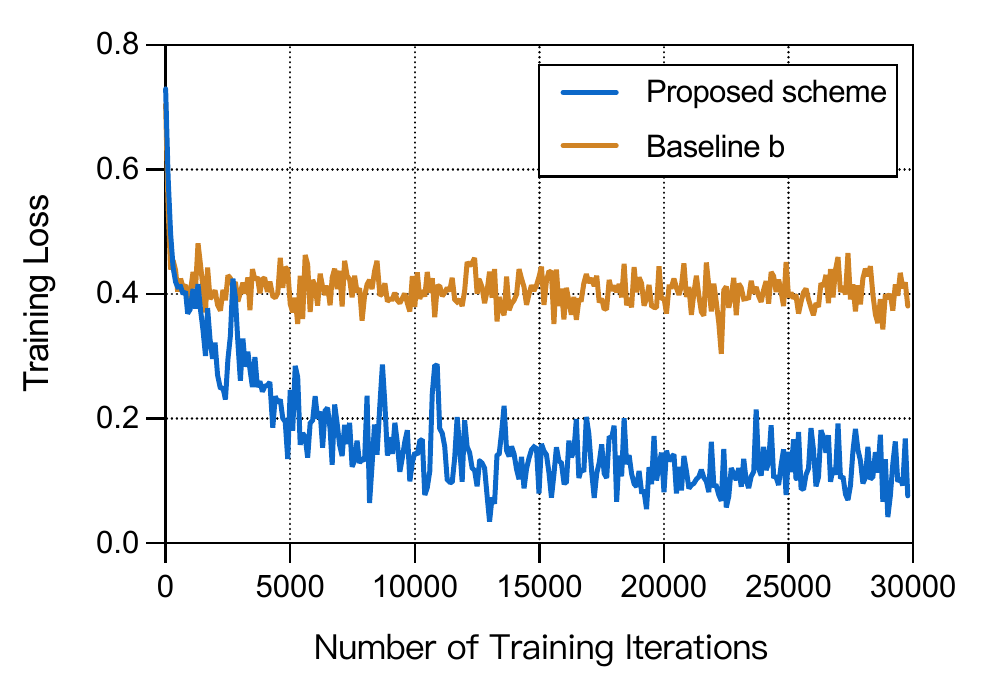}
\caption{\normalsize The training loss as the number of training iterations varies.}
\label{loss}
\end{figure}
In Fig.~\ref{loss}, we show how the training loss changes as the number of training iterations varies. From Fig.~\ref{loss}, we see that, as the number of training iterations increases, the training loss of all considered learning algorithms decreases first and, then remains unchanged. The fact that the training loss remains unchanged demonstrates the convergence of the GNN-based algorithm. From Fig.~\ref{loss}, we can also see that the proposed scheme can reduce training loss by 43.17\%, compared to baseline b. This is due to the fact that the proposed scheme uses different weight matrices (e.g., $\boldsymbol{w}_2$, $\boldsymbol{w}_3$ and $\boldsymbol{w}_4$) for different type of vehicles, and hence, better graph information vectors can be learned.

\begin{figure}[t]
\centering
\includegraphics[width=12cm]{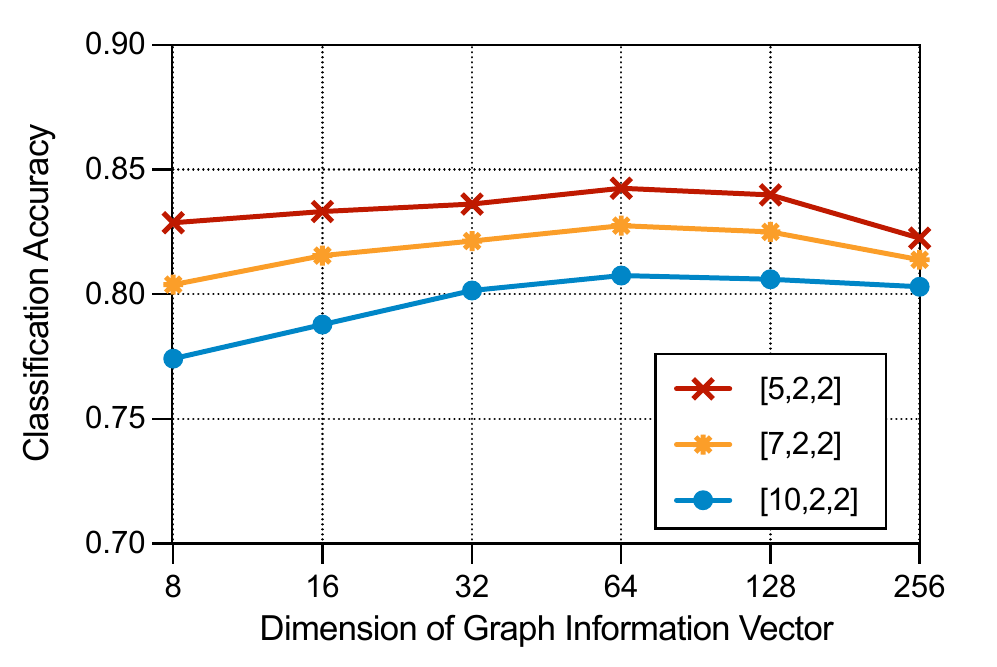}
\caption{\normalsize Impact of the graph information vector dimension $\lambda_0$.}
\label{EmbSize}
\end{figure}
Fig.~\ref{EmbSize} shows how the classification accuracy changes as the dimension of graph information vector varies. 
From Fig.~\ref{EmbSize}, we can see that, as the dimension of graph information vector increases, the classification accuracy increases since better heterogeneous graph representation can be learned. However, as the dimension of graph information vector continues to increase, the performance of all considered algorithms are stabilized. For example, in Fig.~\ref{EmbSize}, all considered algorithms reach the best performance when the dimension of graph information vector $\lambda_0$ is 64, and then the performance becomes stable or even slightly worse. This is due to the fact that, the trained GNN model will be over fitted when the dimension of graph information vector becomes too large.

\begin{figure}[t]
\centering
\subfloat[]{
\begin{minipage}[t]{0.33\linewidth}
\centering
\includegraphics[width=6cm]{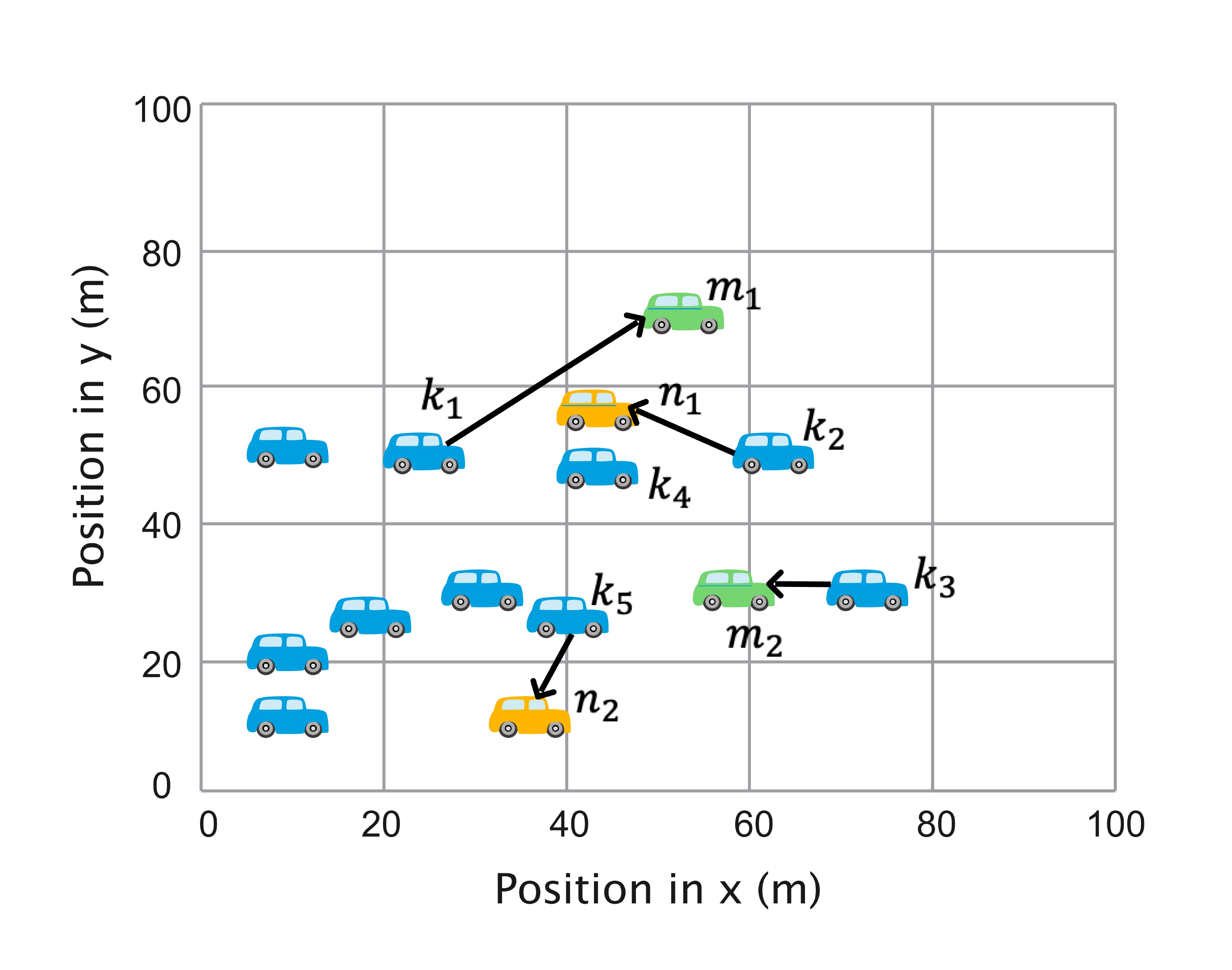}
\label{12a}
\end{minipage}%
}%
\subfloat[]{
\begin{minipage}[t]{0.33\linewidth}
\centering
\includegraphics[width=6cm]{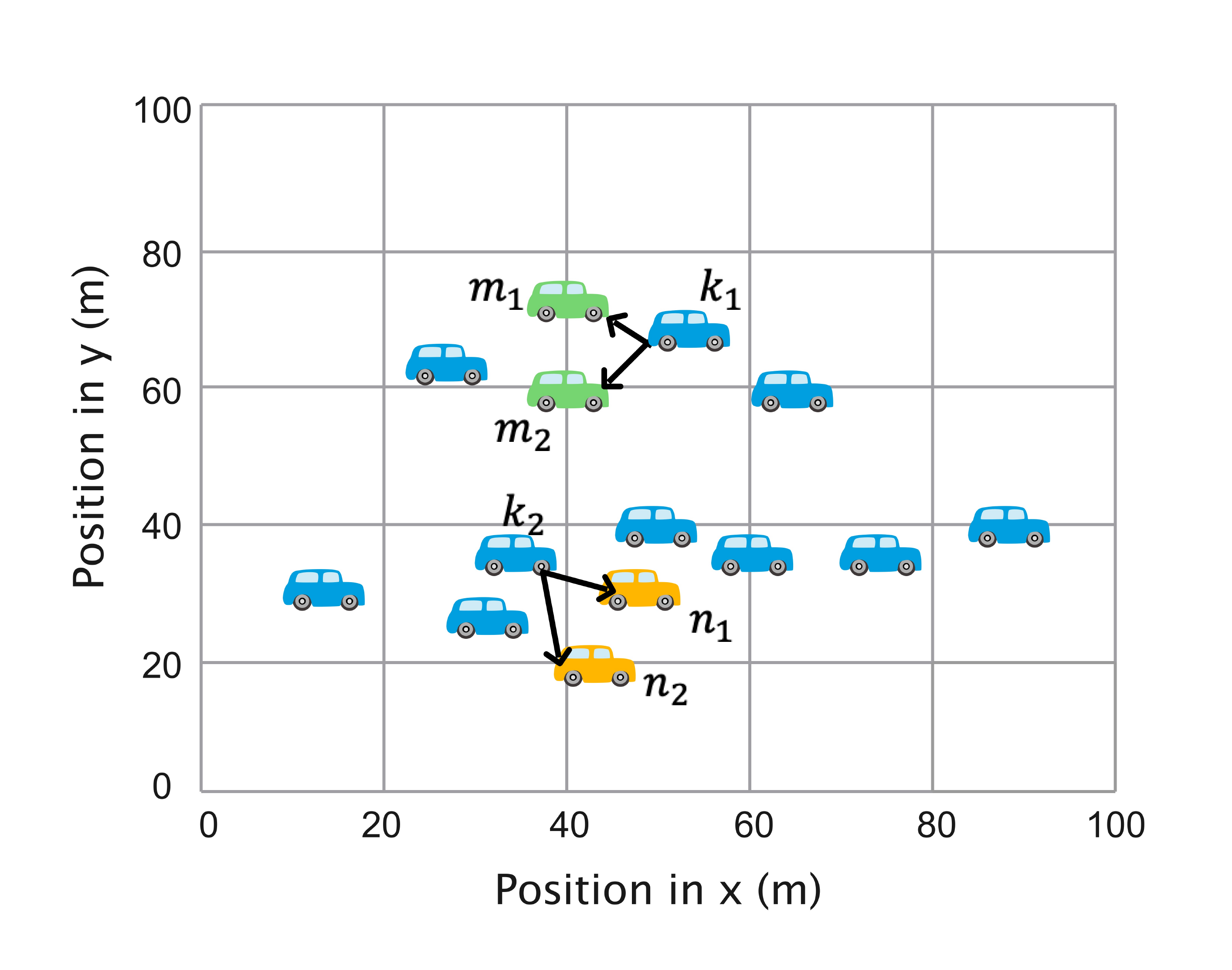}
\label{12b}
\end{minipage}%
}%
\subfloat[]{
\begin{minipage}[t]{0.33\linewidth}
\centering
\includegraphics[width=6cm]{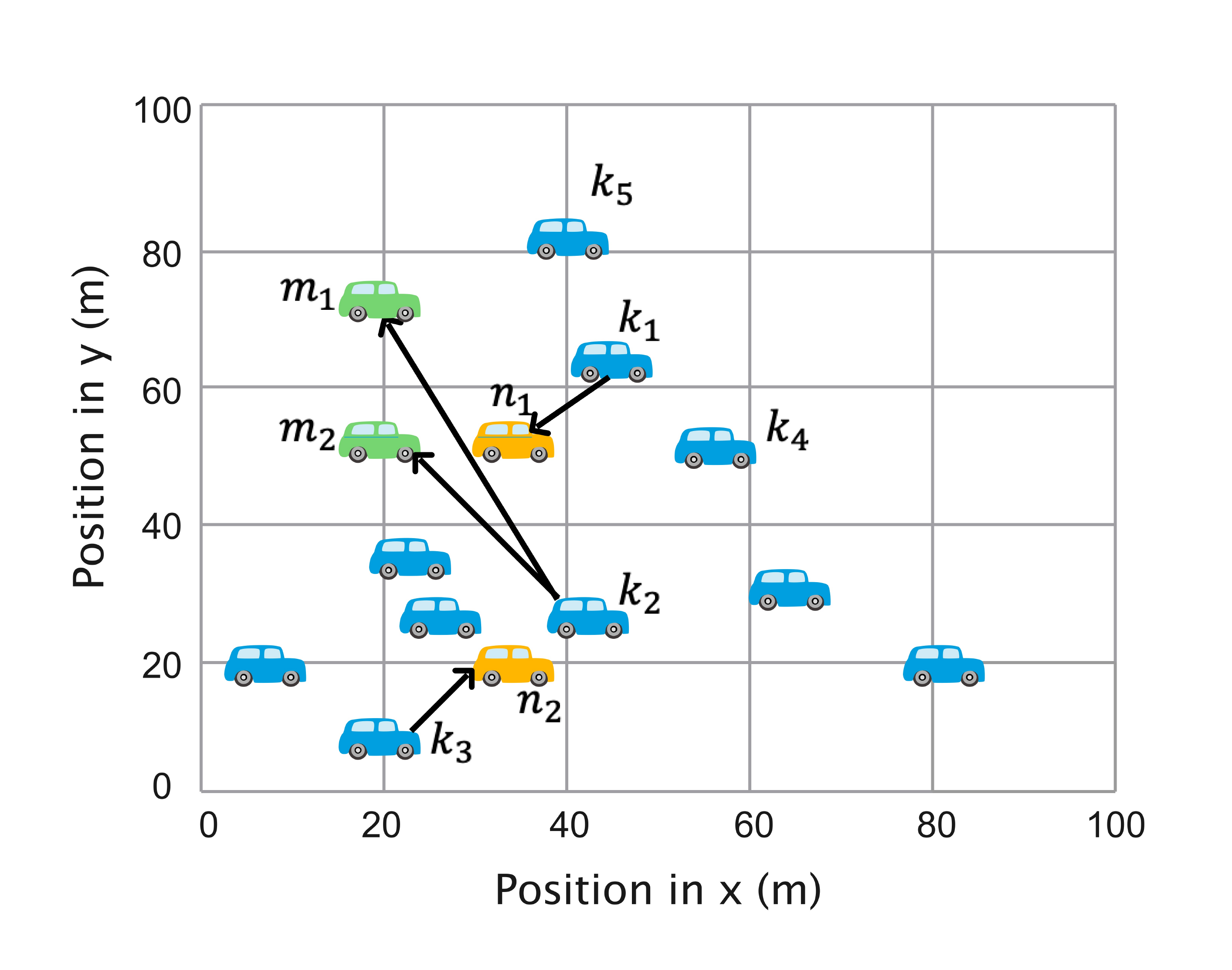}
\label{12c}
\end{minipage}%
}%
\centering
\caption{Visualization of service mode selection and target vehicle association strategy obtained by the proposed method.}
\label{Association}
\end{figure}

Fig.~\ref{Association} is a visualization of using our proposed method for determining the mode of service provider vehicles and target vehicle association. In this figure, the blue, yellow, and green points refer to SPVs, sensing target vehicles, and communication target vehicles, respectively. From Fig.~\ref{Association}\subref{12a}, we can see that the target vehicles do not necessarily select the geographically closest SPV. For example, in Fig.~\ref{Association}\subref{12a}, sensing target vehicle $n_1$ selects SPV $k_2$ instead of its geographically closest SPV $k_4$. This is because the sensing link between vehicle $k_4$ and vehicle $n_1$ will cause interference to the communication link between vehicle $k_1$ and vehicle $m_1$. 
From Fig.~\ref{Association}\subref{12b}, we can also see that if the geographical locations of two target vehicles are close to each other, an SPV is more likely to provide services for both of them simultaneously. For example, SPV $k_1$ provides communication services to communication target vehicles $m_1$ and $m_2$ at the same time. Fig.~\ref{Association}\subref{12c} shows that a sensing target vehicle prefers to select an SPV which can provide a sensing link that is orthogonal to communication links. For example, sensing target vehicle $n_1$ selects SPV $k_1$ instead of SPVs $k_4$, $k_5$. This is because the directions of the communication link $k_2 \rightarrow m_1$ and the sensing link $k_1 \rightarrow n_1$ are nearly orthogonal and hence, the interference between communication link $k_2 \rightarrow m_1$ and sensing link $k_1 \rightarrow n_1$ can be minimized.


\section{Conclusion}
In this paper, we developed a novel framework that uses THz for joint communication and sensing in vehicular networks. Our goal was to maximize the sum of data rates of all communication target vehicles while satisfying the sensing service requirements of all sensing target vehicles. To this end, we formulated an optimization problem that jointly considers service mode selection, target vehicle association, THz channel features, and dynamic vehicle topologies. To solve this problem, we developed a novel heterogeneous GNN-based scheme, which can effectively find the strategy of service mode selection and target vehicle association. The proposed scheme enables the trained GNN to quickly adapt to dynamic vehicle topologies with various vehicle types. Simulation results showed that, compared with the baseline methods, the proposed method can achieve significant gains in terms of the sum rate of all communication target vehicles.

\bibliographystyle{IEEEtran}
\bibliography{7-Jref}

\end{document}